\documentclass [12pt]{article}
\usepackage{lscape}                                           %
\usepackage[centertags]{amsmath}
\usepackage{amsfonts} \usepackage{amssymb}\usepackage{eufrak}
\usepackage{color}
\usepackage{graphicx}
\usepackage{lscape}
\newcommand{\barX}{{\bar X}}

\newcommand{\zb}{{\bar z}}
\newcommand{\bu}{{\bar u}}
\newcommand{\ub}{{\bar u}}

\newcommand{\vb}{{\bar v}}
\newcommand{\bI}{{\bar I}}
\newcommand{\bJ}{{\bar J}}
\newcommand{\bS}{{\bar S}}
\newcommand{\bX}{{\bar X}}

\newcommand{\cD}{{\cal D}}

\newcommand{\cI}{{\cal I}}

\newcommand{\cM}{{\cal M}}
\newcommand{\cN}{{\cal N}}
\newcommand{\cR}{{\cal R}}

\newcommand{\cX}{{\cal X}}
\newcommand{\cbX}{{\bar{\cal X}}}

\newcommand{\Z}{\mathbb{Z}}

\newcommand{\R}{\mathbb{R}}

\newcommand{\C}{\mathbb{C}}

\newcommand{\du}{\partial}
\newcommand{\dub}{{\bar\partial}}
\newcommand{\dz}{\partial}
\newcommand{\dzb}{{\bar\partial}}
\newcommand{\dw}{\partial}

\newcommand{\oh}{\frac{1}{2}}

\newcommand{\COMMENTO}[1]{}
\newcommand{\COMMENTOO}[1]{}
\newcommand{\COMMENTOOK}[1]{}
\newcommand{\COMMENTONO}[1]{}
\begin{document}
%%%%%%%%%%%%%%%%%%%%%%%%%%%%%%%%%%%%%%%%%%%%%%%%%%%%%%%%%%%%%%%%%%%%%%%%%
\title{
Green functions and twist correlators for $N$ branes at angles.
}

\author{%\parbox{11.5cm}
{Igor Pesando$^1$}
\\
~\\
~\\
$^1$Dipartimento di Fisica, Universit\`a di Torino\\
and I.N.F.N. - sezione di Torino \\
Via P. Giuria 1, I-10125 Torino, Italy\\
\vspace{0.3cm}
\\{ipesando@to.infn.it}
}

\maketitle
\thispagestyle{empty}

\abstract{
We compute the Green functions and correlator functions for $N$ twist fields
for branes at angles on $T^2$ and we show that there are $N-2$ %$[N/2]$ 
different configurations labeled by an integer $M$ which is roughly associated
with the number of reflex angles of the configuration.
In order to perform this computation we use a $SL(2,\R)$ 
invariant formulation and geometric constraints instead of Pochammer contours. 
In particular the $M=1$ or $M=N-1$ amplitude can be expressed without using
transcendental functions.
We determine the amplitudes normalization from $N \rightarrow N-1$
reduction without using the factorization into the untwisted sector.
Both the amplitudes normalization and the OPE of two twist fields 
are unique (up to one constant) when  
the $\epsilon \leftrightarrow 1-\epsilon$ symmetry is imposed.
For consistency we find also an infinite number of relations among
Lauricella hypergeometric functions.
}
\\
\\
keywords: {D-branes, Conformal Field Theory}
\\
\\
preprint: {DFTT-6-2012}

\newpage

\section{Introduction and conclusions}

Since the beginning, D-branes have been very important in the formal
development of string theory as well as in attempts to apply string
theory to particle phenomenology and cosmology. 
However, the requirement of chirality in any physically realistic
model  leads to a somewhat restricted number of possible D-brane
set-ups.  An important class are intersecting brane models where
chiral fermions can arise at the intersection of two branes at angles.
An important issue for these models is the computation of Yukawa
couplings and flavour changing neutral currents.

Besides the previous computations many  other computations 
often involve correlators of twist fields and excited
twist fields. 
It is therefore important and interesting in its own to be able to
compute these correlators.
As known in the literature \cite{Dixon:1986qv} 
and explicitly shown in \cite{Pesando:2011ce}
for the case of magnetized branes these computations boil down to the
knowledge of the Green function in presence of twist fields and of the
correlators of the plain twist fields.

In this technical paper we have analyzed the $N$ twist fields amplitudes at tree
level for open strings localized at $D$-branes intersections on $T^2$
using the classical path integral approach \cite{Dixon:1986qv}.
The subject has been explored in many papers and in both the branes at
angles setup and the magnetic branes setup see for example
(\cite{Bianchi:1991rd}, \cite{Gava:1997jt}, \cite{David:2000um},
\cite{Abel:2003yx}, \cite{Cvetic:2003ch}, \cite{Abel:2003vv},
\cite{Bertolini:2005qh}, \cite{Duo:2007he}).

\begin{figure}[hbt]
\begin{center}
\def\svgwidth{250px}
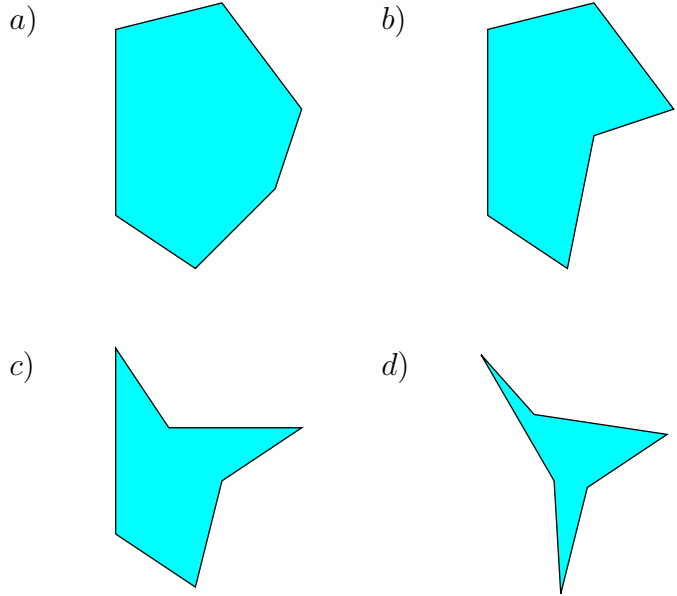
\end{center}
\vskip -0.5cm
\caption{The four different cases with $N=6$. 
$a)$ $M_{ccw}=2$ and $M_{cw}=4$ where $M_{ccw}$ is measured
  counterclockwise and $M_{cw}$ clockwise.
$b)$ $M_{ccw}=3$ and $M_{cw}=3$.
$c)$ $M_{ccw}=4$ and $M_{cw}=2$.
$d)$ $M_{ccw}=5$ and $M_{cw}=1$.
}
\label{fig:6gons}
\end{figure}

We have shown that there are different sectors with different
amplitudes.
Sectors are  labeled by an integer $M_{cw}$ ($1\le M_{cw} \le N-2$)  
and that the number of sectors is equal to the number of reflex angles
formed by the brane configuration. This means that for example all the
configurations in fig. (\ref{fig:6gons}) have different amplitudes.
In particular the quantum amplitudes with $M_{cw}=1$ can be expressed using
elementary functions only. 
This result generalizes the result previously obtained for both four
point amplitudes (\cite{Cvetic:2003ch}, \cite{Abel:2003vv}) and for
the $N$ point amplitudes \cite{Abel:2003yx}
where only the special case $M=N-2$ were considered.
Since the $N=4$ $M=2$ amplitude has also been obtained by a different
approach in (\cite{Anastasopoulos:2011gn}, \cite{Anastasopoulos:2011hj}). 
it would be interesting to understand how this can come about in this
different setup.

We have also obtained the normalizations (up to one constant) 
of both two twist fields OPE and amplitudes.
This result has been achieved using three ingredients: 
the consistency of 
the $N$ twist fields correlator factorization into $N-1$ twist fields
one, 
the canonical normalization of the 2 twist correlator
$\langle \sigma_{\epsilon}(x) \sigma_{1-\epsilon}(y) \rangle
=
1/(x-y)^{\epsilon(1-\epsilon)}
$
%(\ref{2-twist-norm}) 
and the
assumption of the symmetry of under $\sigma_\epsilon \leftrightarrow
\sigma_{1-\epsilon}$. 

Finally we have computed the Green functions in presence of $N$ twist
fields and we have shown that in order to do so there needs three
different kinds of derivatives instead of the usual two which are
needed in the closed string case.

This paper is organized as follows. In section 2 we review the
geometrical framework of branes at angles and we fix our conventions.
In this section we discuss carefully how to make use of the doubling
trick in presence of multiple cuts and the existence of local and
global constraints.
In section 3 we show the existence of $N-2$ different sectors and
compute the corresponding classical solutions. 
We show also explicitly the results for the $N=3$ and $N=4$ cases. 
Moreover using the known
relation between closed string and open string amplitudes
\cite{Kawai:1985xq} we express
the classical action as a sum of products of  holomorphic and
antiholomorphic parts. Details on this computation are given in
appendix \ref{app:KLT}.
In section 4 we compute the Green functions for the different sectors
and give explicit expressions for $N=3$ and $N=4$ cases.
In particular we discuss the existence of infinite relations among
polynomial of  Lauricella hypergeometric functions which must follow from
the consistency of the procedure.
Finally in section 5 we compute the quantum correlators of $N$ twists and their
normalization factors.
In particular we show that the $M_{cw}=1$ sector amplitudes can be
expressed as a  product of elementary functions.
Moreover we discuss how $N-1$ twist fields amplitudes can be obtained
from $N$ twist fields ones. A mathematically curious consequence is
that certain determinant of order $N-2$ involving Lauricella
hypergeometric functions of order $N-3$ are  expressible as product of powers.

\section{Review of branes at angles}
\begin{figure}[hbt]
\begin{center}
\def\svgwidth{350px}
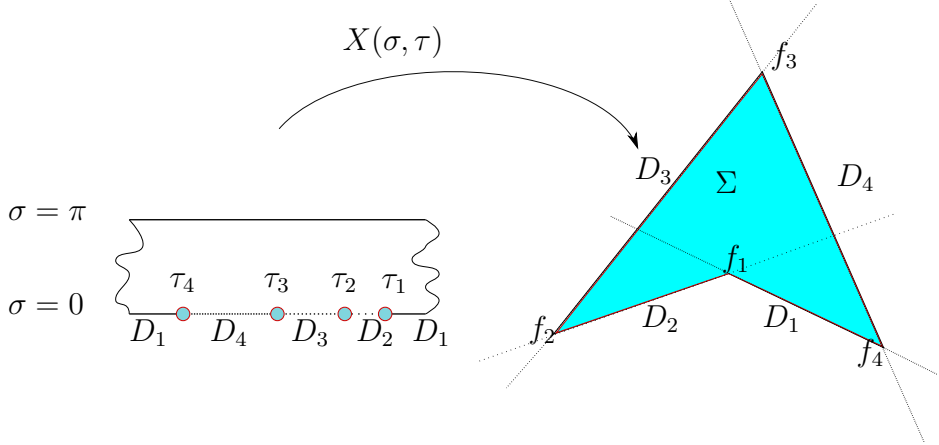
\end{center}
\vskip -0.5cm
\caption{Map from the Minkowskian worldsheet to the target polygon $\Sigma$.}
\label{fig:stripe2polygon}
\end{figure}

The Euclidean action for the string configuration is given by
\begin{equation}
S
=
\frac{1}{4\pi \alpha'} \int d\tau_E \int_0^\pi d\sigma~ (\partial_\alpha X^I)^2
=
\frac{1}{4\pi \alpha'} \int_H d^2 u~ 
(\du X \dub \barX + \dub X \du \barX )
\end{equation}
where $u\in H$, the upper half plane, 
$d^2 u= e^{2 \tau_E} d\tau_E d\sigma= \frac{d u ~d\bar  u}{2 i}$
and $I=1,2$ so that $X= \frac{1}{\sqrt{2}}(X^1+i X^2)$, 
$\barX=X^*$.
The complex string coordinate is a map from the upper half plane to a
closed polygon $\Sigma$ in $\C$, i.e. $X:H \rightarrow \Sigma\subset \C$.
For example in fig. \ref{fig:stripe2polygon} we have pictured the interaction
of $N=4$ branes at angles $D_i$ with $i=1,\dots N$. The interaction
between brane $D_i$ and $D_{i+1}$ is at $f_i\in\C$ where we use the rule
that index $i$ is defined modulo $N$.

\subsection{The local description}
Locally at the interaction point $f_i$
the boundary conditions for the brane $D_{i}$ are given by
\begin{align}
Re( e^{-i \pi \alpha_{i}} X'_{loc} |_{\sigma=0} ) 
=
Im( e^{-i \pi \alpha_{i}} X_{loc} |_{\sigma=0}) -g_{i} 
=0
\label{local-boundary-condition-s=0}
\end{align}
while those for the brane $D_{i+1}$ by
\begin{align}
Re( e^{-i \pi \alpha_{i+1}} X'_{loc} |_{\sigma=\pi} ) 
=
Im( e^{-i \pi \alpha_{i+1}} X_{loc} |_{\sigma=\pi}) -g _{i+1}
=0
\label{local-boundary-condition-s=pi}
\end{align}
with
\begin{equation}
f_i= \frac{e^{i \pi \alpha_{i+1}} g_i-e^{i \pi \alpha_{i}} g_{i+1} }
{\sin~ \pi(\alpha_{i+1}-\alpha_{i})}
\end{equation}

\begin{figure}[hbt]
\begin{center}
\def\svgwidth{300px}
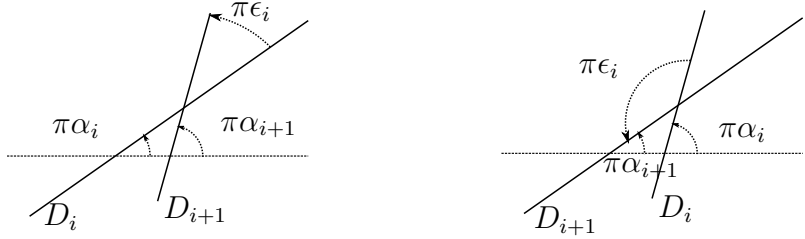
\end{center}
\vskip -0.5cm
\caption{The connection between $\epsilon$ and the geometrical angles
  $\alpha$s defining the branes.}
\label{fig:angles}
\end{figure}

\begin{figure}[hbt]
  \begin{minipage}[t]{0.45\linewidth}
     \def\svgwidth{200px}
     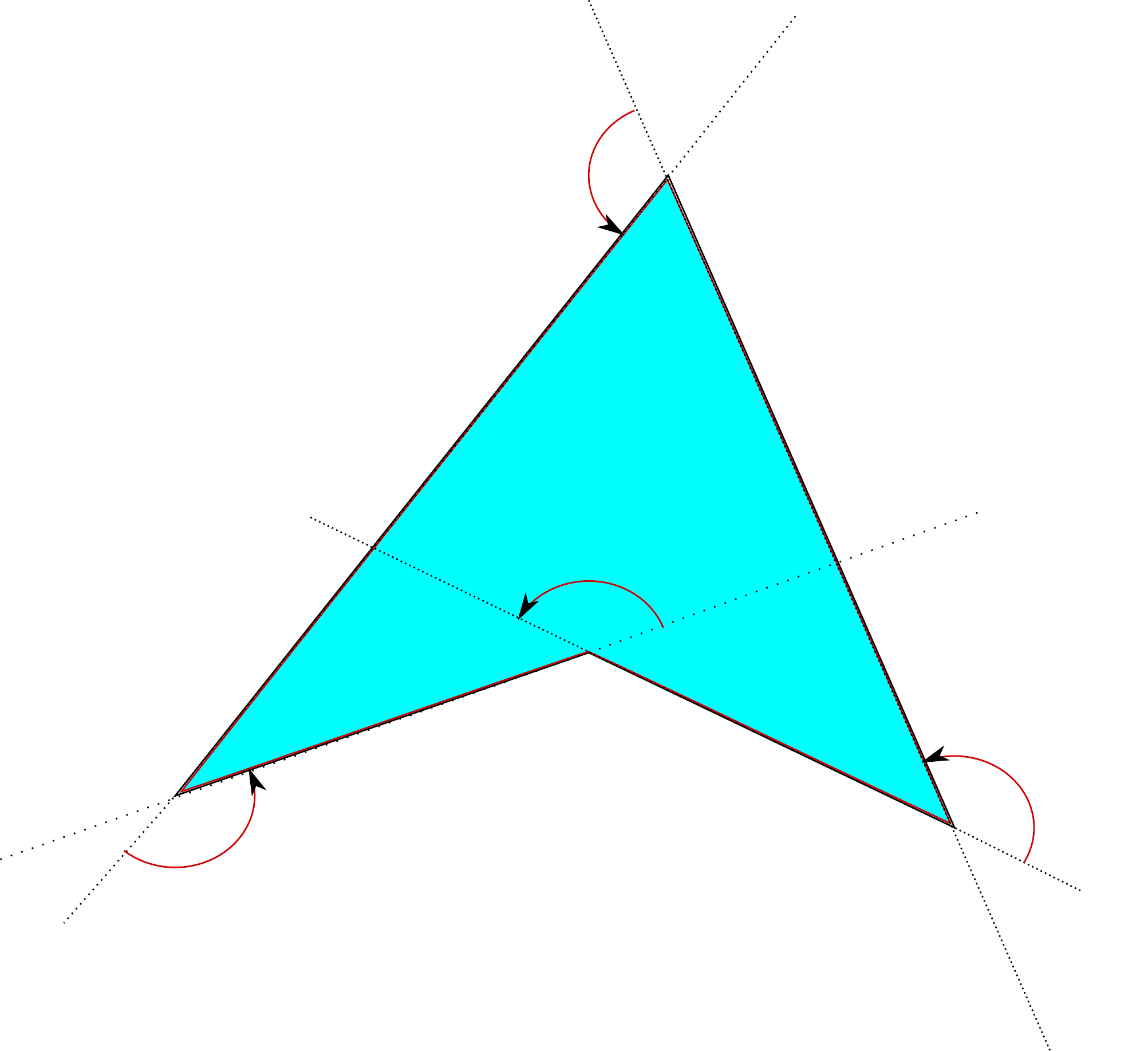
     \vskip -0.5cm
     \caption{A polygon $\Sigma$ with an reflex angle and branes counted
       counterclockwise with $N=4$ and $M_{ccw}=3$. }
     \label{fig:N4Sigma}
  \end{minipage}
  \begin{minipage}[t]{0.45\linewidth}
     \def\svgwidth{200px}
     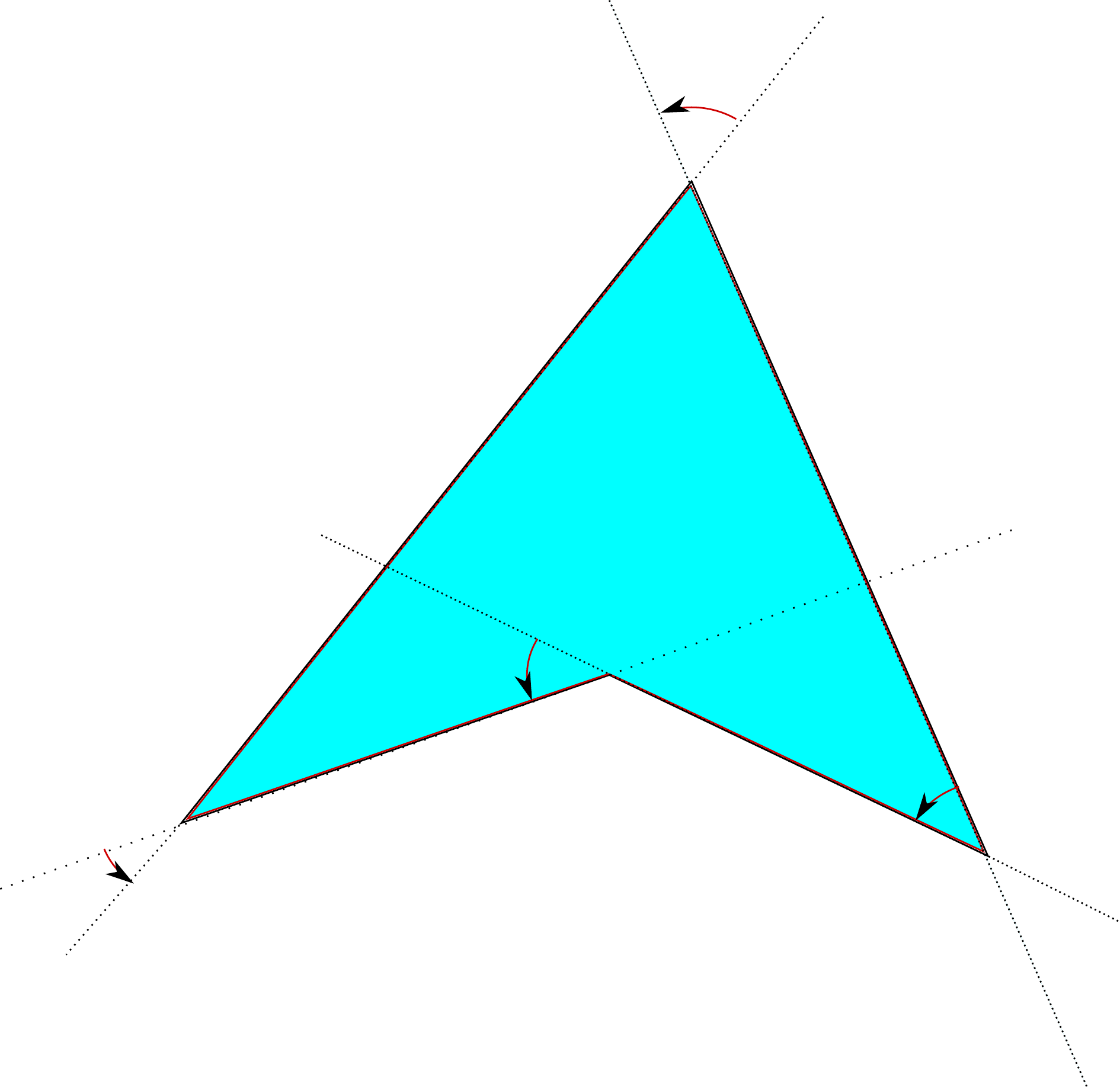
     \vskip -0.5cm
     \caption{A polygon $\Sigma$ with an reflex angle and branes
       counted clockwise with $N=4$ and $M_{cw}=1$. }
     \label{fig:N4Sigma_clockwise}
  \end{minipage}
\end{figure}

When we write the Minkowskian string expansion as 
$X(\sigma,\tau)=X_L(\tau+\sigma)+X_R(\tau-\sigma)$
the previous boundary conditions imply (and not become since they are
not completely equivalent because of zero modes)
\begin{align}
X'_{L~loc}(\xi)= e^{i 2\pi \alpha_i} X'_{R~loc}(\xi)
,~~~~
X'_{L~loc}(\xi+\pi)= e^{i 2\pi \alpha_{i+1}} X'_{R~loc}(\xi-\pi)
\label{local-boundary-condition-upper}
\end{align}
or in a more useful way in order to explicitly compute the mode expansion
\begin{align}
X'_{L~loc}(\xi+2\pi)= e^{i 2\pi \epsilon_i} X'_{L~loc}(\xi)
,~~~~
X'_{R~loc}(\xi+2\pi)= e^{-i 2\pi \epsilon_i} X'_{R~loc}(\xi)
\label{local-boundary-condition}
\end{align}
where we have defined
\begin{equation}
\epsilon_i=
\left\{\begin{array}{c c}
(\alpha_{i+1}-\alpha_i) & \alpha_{i+1}>\alpha_i
\\
1+(\alpha_{i+1}-\alpha_i) & \alpha_{i+1}<\alpha_i
\end{array}
\right.
\label{eps-alf-alf}
\end{equation}
so that $0<\epsilon_i<1$ and there is no ambiguity in the phase 
$e^{i  2\pi \epsilon_i}$ entering the boundary conditions.
The quantity $\pi \epsilon_i$ is the angle between the two branes
$D_i$ and $D_{i+1}$ measured counterclockwise as shown in
fig. \ref{fig:angles}. 
A consequence of this definition is that $\epsilon$ becomes
$1-\epsilon$  when we flip the order of two branes.
For example the angles in fig. \ref{fig:N4Sigma} become those in
fig. \ref{fig:N4Sigma_clockwise} when we reverse the order we count
the branes, i.e. when we follow the boundary clockwise instead of
counterclockwise  the physics must obviously not change.

We introduce as usual the Euclidean fields 
$X_{loc}(u,\bar u)$, $\bar X_{loc}(u,\bar u)$ 
by a worldsheet Wick rotation in such a way they are defined on the 
upper half plane by $u=e^{\tau_E+i\sigma} \in H$. 
The previous choice of having brane $D_i$ at $\sigma=0$
(\ref{local-boundary-condition-s=0}) and brane $D_{i+1}$ at
$\sigma=\pi$ (\ref{local-boundary-condition-s=pi}) implies that 
in the local description where the interaction point is at $x=0$ $D_i$
is mapped into $x>0$ and $D_{i+1}$ into $x<0$.
The boundary conditions (\ref{local-boundary-condition-upper}) can
then immediately be written as
\begin{align}
 \partial X_{loc}(x+ i 0^+ ) 
&= 
e^{i 2 \pi \alpha_i}
 \bar \partial \bar X_{loc}(x - i 0^+)
~~
0<x,
\nonumber\\
%~~~~
 \partial X_{loc}(x+ i 0^+ ) 
&= 
e^{i 2 \pi \alpha_{i+1}}
\bar  \partial \bar X_{loc}(x - i 0^+)
~~~~
x<0
\label{local-boundary-upper}
\end{align}
and similarly relations for $\bX$ which can be obtained by complex conjugation. 
When we add to the previous conditions the further constraints
\begin{equation}
X(0,0)= f_i,~~~~
\bar X(0,0)= f_i^*
\label{local-boundary-points}
\end{equation}
we obtain a system of conditions which are equivalent to the original
ones (\ref{local-boundary-condition-s=0}, \ref{local-boundary-condition-s=pi}).

In order to express the boundary conditions 
(\ref{local-boundary-condition}) in the Euclidean formulation
it is better to introduce the local fields 
defined on the whole complex plane by the doubling trick as
\begin{align}
\dz\cX_{loc}(z)
&=
\left\{
\begin{array}{cc}
  \du X_{loc}(u) 
  & z=u\mbox{ with }{Im~} z >0\mbox{ or } z\in\R^+ 
  \\
  e^{i 2\pi \alpha_i} \dub \bX_{loc}(\bar u) 
  & z=\bar u\mbox{ with }{Im~} z <0\mbox{ or } z\in\R^+ 
\end{array}
\right.
\nonumber\\
\dz \cbX_{loc}(z)
&=
\left\{
\begin{array}{cc}
  \du \bX_{loc}(u) 
  & 
  z= u\mbox{ with }{Im~} z >0\mbox{ or } z\in\R^+ 
  \\
  e^{-i 2\pi \alpha_i} \dub X_{loc}(\bar u) 
  & 
  z=\bar u\mbox{ with }{Im~} z <0\mbox{ or } z\in\R^+ 
\end{array}
\right.
\label{loc-calX-calXbar}
\end{align}
In this way we can write  eq.s (\ref{local-boundary-condition}) as  
\begin{align}
 \partial\cX_{loc}(e^{i 2\pi} \delta ) 
= 
e^{i 2 \pi \epsilon_i}
 \partial\cX_{loc}(\delta)
%\nonumber\\
,~~~~
 \partial\cbX_{loc}(e^{i 2\pi} \delta ) 
= 
e^{-i 2 \pi \epsilon_i}
 \partial\cbX_{loc}(\delta)
\end{align}
%when we place the cut on the real axis toward $-\infty$.
Notice that while the two Minkowskian boundary conditions
(\ref{local-boundary-condition}) are one the complex conjugate of
the other the previous Euclidean ones are independent and each is
mapped into itself by complex conjugation therefore the Euclidean
classical solutions for $\cX$ and $\cbX$ are independent.

The quantization of the string with given boundary conditions yields
\begin{align}
X_{loc}(u,\bar u)
&=
f_i
&+
i\oh \sqrt{2\alpha'}
e^{i \pi \alpha_{i}}
\sum_{n=0}^\infty 
\left[
\frac{\bar \alpha_{(i)n}}{ n+1-\epsilon_i} u^{-(n+1-\epsilon_i)}
-
\frac{\alpha_{(i)n}^\dagger}{ n+\epsilon_i} u^{n+\epsilon_i}
\right]
\nonumber\\
&
&+
i \oh \sqrt{2\alpha'}
e^{i \pi \alpha_{i}}
\sum_{n=0}^\infty
\left[ 
-
\frac{\bar \alpha_{(i)n}^\dagger}{ n+1-\epsilon_i} \bar u^{n+1-\epsilon_i}
+
\frac{\alpha_{(i)n}}{ n+\epsilon_i} \bar u^{-(n+\epsilon_i)}
\right]
\nonumber\\
\bar X_{loc}(u,\bar u)
&=
f_i^*
&+
i \oh \sqrt{2\alpha'}
e^{-i \pi \alpha_{i}}
\sum_{n=0}^\infty 
\left[
-
\frac{\bar \alpha_{(i)n}^\dagger}{ n+1-\epsilon_i} u^{n+1-\epsilon_i}
+
\frac{\alpha_{(i)n}}{ n+\epsilon_i} u^{-(n+\epsilon_i)}
\right]
\nonumber\\
&
&+
i \oh \sqrt{2\alpha'}
e^{-i \pi \alpha_{i}}
\sum_{n=0}^\infty 
\left[
\frac{\bar \alpha_{(i)n}}{ n+1-\epsilon_i} \bar u^{-(n+1-\epsilon_i)}
-
\frac{\alpha_{(i)n}^\dagger}{ n+\epsilon_i} \bar u^{n+\epsilon_i}
\right]
\end{align}
with non trivial commutation relations ($n,m \ge 0$)
\begin{equation}
[ \alpha_{(i) n}, \alpha_{(i) m}^\dagger]
=( n+\epsilon_i) \delta_{m,n}
,~~~~
[ \bar\alpha_{(i) n}, \bar\alpha_{(i) m}^\dagger]
=( n+1-\epsilon_i) \delta_{m,n}
\end{equation}
and vacuum defined in the usual way by
\begin{equation}
\alpha_{(i) n} |T_i\rangle =\bar \alpha_{(i) n }|T_i\rangle =0
~~~~ n\ge 0
\end{equation}
The vacuum is then generated from the twist operator
$\sigma_{\epsilon_i,f_i}$ which depends both on the twist $\epsilon_i$
and  on the position $f_i\in\C$.
The dependence on the twist $\epsilon_i$ can be read f.x. from the OPEs
\begin{align}
\du X(u) \sigma_{\epsilon_i,f_i}(x) 
&\sim (u-x)^{\epsilon_i-1} (\du X \sigma_{\epsilon_i,f_i})(x)
\nonumber\\
\du \bX(u) \sigma_{\epsilon_i,f_i}(x) &\sim (u-x)^{-\epsilon_i} 
(\du \bX \sigma_{\epsilon_i,f_i})(x)
\label{OPEs-epsilon}
\end{align}
which can be deduced from the local computations
\begin{align}
\du X_{loc}(u) |T_i\rangle \sim  
u^{\epsilon_i-1} ~(-i\oh\sqrt{2\alpha'} e^{i \pi \alpha_{i-1}} \alpha_{(i)0}^\dagger |T_i\rangle )
,~~~~
\du \bX_{loc}(u) |T_i\rangle \sim
u^{-\epsilon_i} ~(-i\oh\sqrt{2\alpha'} e^{-i \pi \alpha_{i-1}} \bar \alpha_{(i)0}^\dagger |T_i\rangle )
\end{align}
On the other side the dependence on $f_i$ can be read from the OPE
\begin{equation}
e^{i k\cdot X(z,\bar z)}  \sigma_{\epsilon_i,f_i}(x)  \sim 
|z|^{ - \alpha' k^2} e^{-\frac{1}{2} R^2(\epsilon_i)  \alpha' k^2 } 
e^{i k \cdot f_i}  \sigma_{\epsilon_i,f_i}(x)
\end{equation}
which can be deduced from the local computation
\begin{equation}
|z|^{ -\alpha' k^2} e^{-\frac{1}{2} R^2(\epsilon_i)  \alpha' k^2 } 
:e^{i k \cdot X_{loc}(z,\bar z)}: |T_i \rangle
\sim |z|^{ -\alpha' k^2} e^{i k \cdot f_i} 
e^{-\frac{1}{2} R^2(\epsilon_i)  \alpha' k^2 } |T_i \rangle
\end{equation}
upon the identification \cite{Pesando:2011yd}
$e^{i k\cdot X(z,\bar z)} \leftrightarrow 
|z|^{ -\alpha' k^2}  e^{-\frac{1}{2} R^2(\epsilon_i)  \alpha' k^2 } 
:e^{i k  \cdot X_{loc}(z,\bar z)}:$
with $R^2(\epsilon_i)=2 \psi(1)- \psi(\epsilon_i)
-\psi(1-\epsilon_i)$,
$\psi(z)= \frac{d \ln \Gamma(z)}{d z}$ being the digamma function. 
Notice that there is no obvious way of computing the angles
$\alpha_{i}$ and $\alpha_{i+1}$ from OPEs.

\subsection{Global description}
In the local description, where the interaction point is at $x=0$, $D_i$
is mapped into $x>0$ and $D_{i+1}$ into $x<0$
this means that in the global description the
world sheet interaction points are mapped on the boundary of the upper
half plane  so that $x_{i+1}< x_i$.
The global equivalent of the local boundary conditions 
eq.s (\ref{local-boundary-upper}) become
\begin{align}
 \partial X_L(x+ i 0^+ ) 
&= 
e^{i 2 \pi \alpha_i}
 \bar \partial \bar X_R(x - i 0^+)
~~~~
x_{i}<x<x_{i-1}
\nonumber\\
 \partial \bX_L(x+ i 0^+ ) 
&= 
e^{-i 2 \pi \alpha_i}
 \bar \partial X_R(x - i 0^+)
~~~~
x_{i}<x<x_{i-1}
\label{global-boundary-upper}
\end{align}
To the previous constraints one must also add
\begin{equation}
X_{loc}(x_i,\bar x_i)= f_i,~~~~
\bar X_{loc}(x_i ,\bar x_i)= f_i^*
\label{global-boundary-points}
\end{equation}
in order to get a system of boundary conditions equivalent to the
original ones (\ref{local-boundary-condition-s=0},
\ref{local-boundary-condition-s=pi}). 
When we introduce the global fields 
defined on the whole complex plane by the doubling trick as\footnote{
It is also possible to perform the doubling trick by defining
\begin{align*}
\dz\cX(z)
&=
\left\{
\begin{array}{cc}
  \du X(u) 
  & z=u\mbox{ with }{Im~} z >0\mbox{ or } z\in\R-[x_N,x_1] 
  \\
  e^{i 2\pi \alpha_1} \dub \bX(\bar u) 
  & z=\bar u\mbox{ with }{Im~} z <0\mbox{ or } z\in\R-[x_N,x_1] 
\end{array}
\right.
\end{align*}
and similarly for $\dz \cbX(z)$ but then all the formulae 
require a cyclic permutation of the indexes as $2 \rightarrow 1
\rightarrow N \rightarrow 1 $ so that the anharmonic ratio (\ref{anha-ratio})
becomes
$ \omega_z= \frac{(z-x_2) (x_1-x_N)}{ (z-x_N) (x_1-x_2)}
$.
This is not a cyclic permutation for all indexes, i.e it is not 
$ i \rightarrow i-1$ and hence 
and  all the $x_{j \ne 1,2,N}$ are mapped to $\omega_j<0$,
nevertheless 
$\sum_{j=2}^N \equiv  \sum_{j\ne 1} \rightarrow \sum_{j\ne N } \equiv
\sum_{j=1}^{N-1}$
where in order to perform the change of indexes we have rewritten
$\sum_{j=2}^N$ as $ \sum_{j\ne 1}$ and similarly for the product
$\prod_{j=2}^N  \rightarrow  \prod_{j=1}^{N-1}$.
There is also a third possibility and amounts to a cyclic permutation 
$2 \rightarrow 1 \rightarrow N \rightarrow N-1 \rightarrow
... \rightarrow 2$.
}
\begin{align}
\dz\cX(z)
&=
\left\{
\begin{array}{cc}
  \du X(u) 
  & z=u\mbox{ with }{Im~} z >0\mbox{ or } z\in\R-[x_N,x_2]-[x_1,\infty] 
  \\
  e^{i 2\pi \alpha_2} \dub \bX(\bar u) 
  & z=\bar u\mbox{ with }{Im~} z <0\mbox{ or } z\in\R-[x_N,x_2]-[x_1,\infty] 
\end{array}
\right.
\nonumber\\
\dz \cbX(z)
&=
\left\{
\begin{array}{cc}
  \du \bX(u) 
  & 
  z= u\mbox{ with }{Im~} z >0\mbox{ or } z\in\R-[x_N,x_2]-[x_1,\infty] 
  \\
  e^{-i 2\pi \alpha_2} \dub X(\bar u) 
  & 
  z=\bar u\mbox{ with }{Im~} z <0\mbox{ or } z\in\R-[x_N,x_2]-[x_1,\infty] 
\end{array}
\right.
\label{calX-calXbar}
\end{align}
the local boundary conditions (\ref{local-boundary-condition}) 
can be written in the global formulation as
\begin{align}
 \partial\cX(x_i+ e^{i 2\pi} \delta ) 
&= 
e^{i 2 \pi \epsilon_i}
 \partial\cX(x_i +\delta)
\nonumber\\
 \partial\cbX(x_i+ e^{i 2\pi} \delta ) 
&= 
e^{-i 2 \pi \epsilon_i}
 \partial\cbX(x_i +\delta)
.
\end{align}
For the proper definition of the global constraints which follow from
eq.s (\ref{global-boundary-points}), for example 
when dealing with the derivatives of the Green functions as in section
\ref{sect:Green_functions}  it is worth noticing the behavior
of the previously introduced fields under complex conjugation when $z$
is restricted to $z\in \C-[-\infty,x_2]-[x_1,\infty]$
\begin{align}
[ \dz \cX(z)]^*
&=
e^{-i 2\pi \alpha _2} \dz \cX(z\rightarrow \bar z) 
=
\dzb \cbX(\bar z)
=
\left\{
\begin{array}{c c}
  \dub \bX(\bu) 
  & \zb=\ub%\mbox{ with }{Im~} z >0\mbox{ or } z\in\R-[x_N,x_1] 
  \\
  e^{-i 2\pi \alpha_2} \du X(u) 
  & \zb= u%\mbox{ with }{Im~} z <0\mbox{ or } z\in\R-[x_N,x_1] 
\end{array}
\right.
\nonumber\\
[ \dz \cbX(z)]^*
&=
e^{-i 2\pi \alpha _2} \dz \cbX(z\rightarrow \bar z) 
=
\dzb \cX(\bar z)
=
\left\{
\begin{array}{c c}
  \dub X(\bu) 
  & \zb=\ub%\mbox{ with }{Im~} z >0\mbox{ or } z\in\R-[x_N,x_1] 
  \\
  e^{i 2\pi \alpha_2} \du \bX(u) 
  & \zb= u%\mbox{ with }{Im~} z <0\mbox{ or } z\in\R-[x_N,x_1] 
\end{array}
\right.
\label{dcX-dbcX}
\end{align}
where $ \dz \cX(z\rightarrow \bar z)$ means that the holomorphic $\dz
\cX(z)$ is evaluated at $\zb$.
The previous expressions also show that it is not necessary to introduce the
antiholomorphic fields $\dzb \cX(\bar z)$ and
 $\dzb \cbX(\bar z)$ which it is possible to construct applying the doubling
trick on  $\dub X(\bar u)$ and $\dub \bX(\bar u)$ respectively.
%%%%%%%%%%%%%%%%%%%%%%%%%%%%%%%%%%%%%%%%%%%%%%%%%%%%%%%%%%%%%%%%%%%%%%
%%%%%%%%%%%%%%%%%%%%%%%%%%%%%%%%%%%%%%%%%%%%%%%%%%%%%%%%%%%%%%%%%%%%%%
%%%%%%%%%%%%%%%%%%%%%%%%%%%%%%%%%%%%%%%%%%%%%%%%%%%%%%%%%%%%%%%%%%%%%%
\section{The path integral approach}
Following the by now classic method \cite{Dixon:1986qv} 
we compute twists correlators by the path integral
\begin{align}
\langle 
\sigma_{\epsilon_1,f_1}(x_1) \dots \sigma_{\epsilon_N,f_N}(x_N)
\rangle
=\int_{\cM(\{x_i,\epsilon_i, f_i\})} \cD X e^{-S_E}
\end{align}
where $\cM(\{x_i, \epsilon_i, f_i\})$ is the space of string configurations
satisfying the boundary conditions (\ref{global-boundary-upper}) and 
(\ref{global-boundary-points}).
Since the integral is quadratic we can then efficiently 
separate the classical fields from the quantum  fluctuations as
\begin{equation}
X(u,\bar u)= X_{cl}(u,\bar u)+X_{q}(u,\bar u)
\end{equation}
where $X_{cl}$ satisfies the previous boundary conditions while $X_q$
satisfies the same  boundary conditions but with all $f_i=0$.
After this splitting we obtain
\begin{align}
\langle 
\sigma_{\epsilon_1,f_1}(x_1) \dots \sigma_{\epsilon_N,f_N}(x_N)
\rangle
=
\cN(x_i,\epsilon_i)
 e^{-S_{E,cl}(x_i,\epsilon_i, f_i)}
\label{general_corr_from_quantum_and_classical}
\end{align}
The explicit expressions for $X_{cl}$ and  $\bar X_{cl}$ given in eq.s
(\ref{Xcl}, \ref{barXcl}) 
show that they vanish when $f_i=0$
hence also the classical action evaluated for $f_i=0$
$S_{E,cl}(x_i,\epsilon_i, f_i=0)$ is zero.
Actually because of translational invariance what said before works
even when all $f_i$ are equal, i.e. when $f_i=f$  and 
therefore we can identify
\begin{equation}
\cN(x_i,\epsilon_i)
=
\langle 
\sigma_{\epsilon_1,f_1=f}(x_1) \dots \sigma_{\epsilon_N,f_N=f}(x_N)
\rangle
\end{equation}
Our strategy is therefore first to compute  the classical contribution
in the rest of this section and then compute the quantum contribution
in section \ref{sect:quantum_corr}.

\subsection{The classical solution}
\begin{figure}[hbt]
\begin{center}
\def\svgwidth{200px}
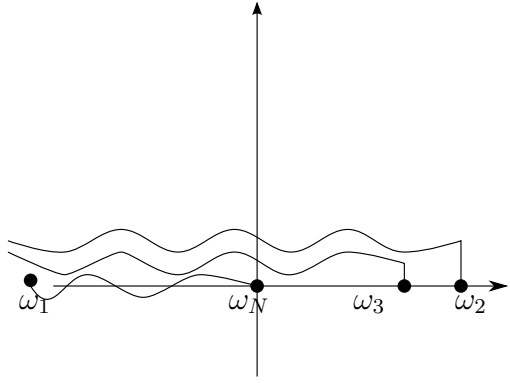
\end{center}
\vskip -0.5cm
\caption{Cuts and prevertexes positions in the $\omega$ plane.}
\label{fig:omega_cuts}
\end{figure}

We want now to write the general solution for $\dz\cX$ and $\dz\cbX$
in a way that  the $SL(2,\R)$ symmetry is manifest.
To this purpose we introduce the anharmonic ratio
\begin{equation}
\omega_z= \frac{(z-x_N) (x_2-x_1)}{ (z-x_1) (x_2-x_N)}
\label{anha-ratio}
\end{equation}
and the corresponding ones $\omega_j$ where $z$ has been replaced with
$x_j$. In particular we get $\omega_N=0$, $\omega_2=1$ and
$\omega_1=-\infty$.
The choice of $\omega_1=-\infty$ is dictated by the request that
powers are defined as $ (\omega-\omega_i)^\epsilon=
|\omega-\omega_i|^\epsilon e^{i \phi \epsilon}$ where
$\phi=arg(\omega-\omega_i)$ is counted from the real axis with range
$(-\pi,\pi)$ so that all cuts must be towards $-\infty$, as it is
shown in fig. (\ref{fig:omega_cuts}) 

\begin{figure}[hbt]
\begin{center}
\def\svgwidth{350px}
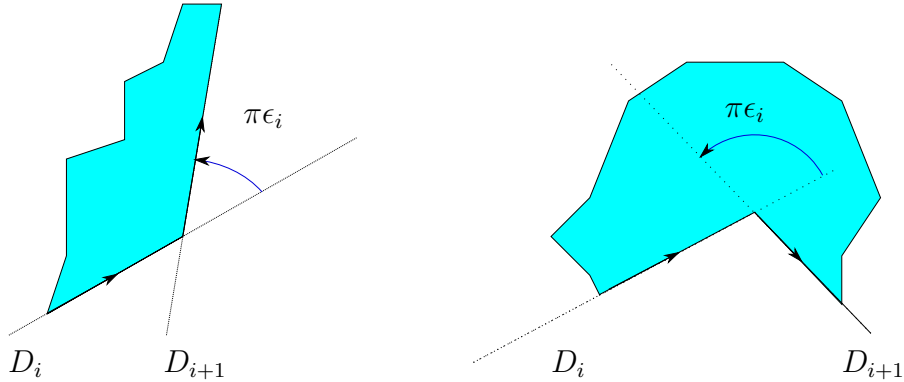
\end{center}
\vskip -0.5cm
\caption{When the target polygon $\Sigma$ (the shaded area) 
is followed counterclockwise keeping the interior on the left side 
an internal acute angle is equal to
$\pi-\pi\epsilon$ while an reflex one  to $2\pi-\pi\epsilon$. }
\label{fig:Acute-Reflex-Contribution}
\end{figure}

\begin{figure}[hbt]
\begin{center}
\def\svgwidth{100px}
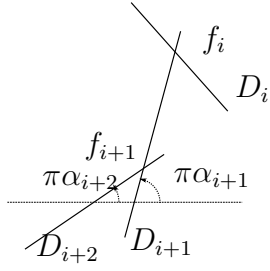
\end{center}
\vskip -0.5cm
\caption{The connection between $f_{i+1}-f_{i}$ and the geometrical angle
  $\alpha_{i+1}$ defining the brane.}
\label{fig:diff_f_angle}
\end{figure}

We can now write the general solutions as
\begin{align}
\dz\cX(z)
&=
\frac{\partial \omega_z}{\partial z}
~\sum_{n=0}^{N-M-2} a_n(\omega_j) \dz_\omega\cX^{(n)}(\omega_z)
\nonumber\\
\dz\cbX(z)
&=
e^{-i 2\pi \alpha_2}
\frac{\partial \omega_z}{\partial z}
~\sum_{r=0}^{M-2} b_r(\omega_j) \dz_\omega\cbX^{(r)}(\omega_z)
\label{class-sols}
\end{align}
where we have defined the basis
\begin{align}
\dz_\omega\cX^{(n)}(\omega_z)
&=
~\prod_{j=2}^N (\omega_z-\omega_j)^{-(1-\epsilon_j)} ~\omega_z^n
,~~~~ 0\le n \le N-M-2
\nonumber\\
\dz_\omega\cbX^{(r)}(\omega_z)
&=
~\prod_{j=2}^N (\omega_z-\omega_j)^{-\epsilon_j}~\omega_z^r
,~~~~ 0\le r \le M-2
\label{basis-class-sols}
\end{align}
and we have also defined the integer
\begin{equation}
M=\sum_{i=1}^N \epsilon_i
\end{equation}
When the target polygon $\Sigma$ is followed counterclockwise 
this integer $M$ is equal to the number of reflex angles plus 2 since
every acute angle internal the target polygon is $\pi-\pi\epsilon$
while every reflex one is $2\pi-\pi\epsilon$
as shown in fig. (\ref {fig:Acute-Reflex-Contribution}).
In a similar way when the target polygon is followed clockwise
$M$ is  the number of acute angles minus 2.
Nevertheless it is important to notice how polygons having $M$ and
$M'=N-M$ both measured counterclockwise or clockwise are not the same
polygons as it shown in fig. (\ref{fig:6gons}) in the case $N=6$.
\COMMENTOOK{ come distinguere se ccw o cw? Non posso se non guardando
  se usando $f$ posso costruire la figura.}
To distinguish  between these two cases it is necessary to compare
expression (\ref{eps-alf-alf}) with 
the phases $\alpha_i$ as derived from the geometrical relations
$ f_{i+1}-f_{i}= \pm e^{i \pi \alpha_{i+1}} |f_{i+1}-f_{i}|$ (where
the sign depends on the case) as shown in
fig.  (\ref{fig:diff_f_angle})

\begin{figure}[hbt]
\begin{center}
\def\svgwidth{400px}
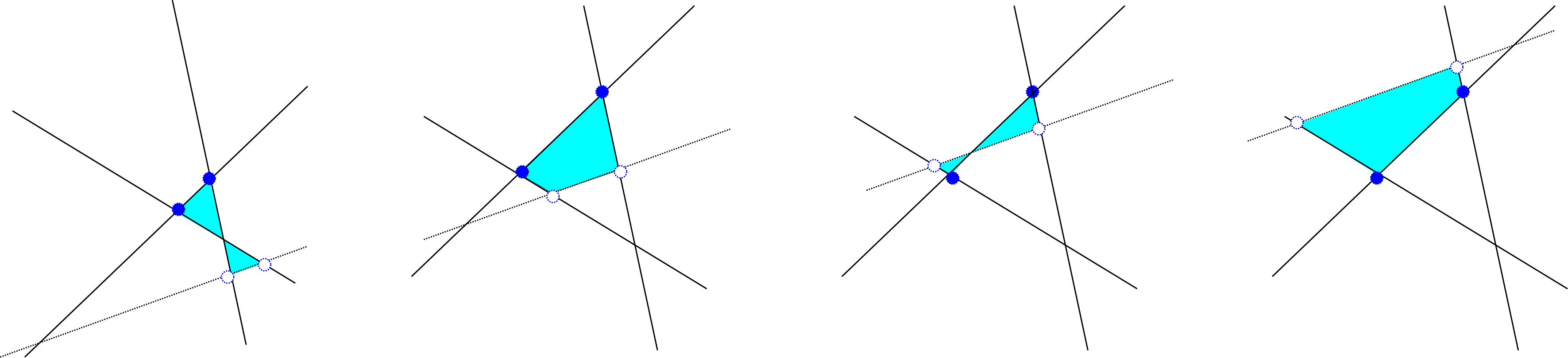
\end{center}
\vskip -0.5cm
\caption{The four different cases with $N=4$ and $M_{ccw}=2$ and  $M_{cw}=2$ 
which can be obtained moving the brane whose intersection points are
the empty circles.
}
\label{fig:change_f_N4M2}
\end{figure}

\begin{figure}[hbt]
\begin{center}
\def\svgwidth{400px}
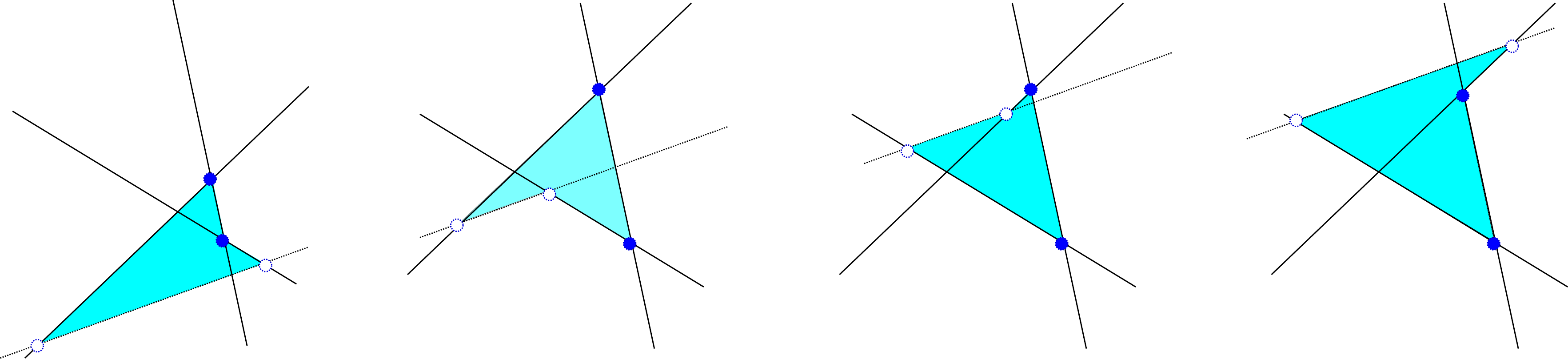
\end{center}
\vskip -0.5cm
\caption{The four different cases with $N=4$ and $M_{ccw}=3$ and  $M_{cw}=1$ 
which can be obtained moving whose intersection points are
the empty circles.
}
\label{fig:change_f_N4M3}
\end{figure}

\COMMENTONO{
Another way of understating this is to consider what happens to 
the constraint $\sum_{i=1}^N \phi_i= (N-2)\pi$  on the internal
geometrical angles $\phi_i$ under the mapping $ \epsilon \rightarrow
1-\epsilon$; the result is that $\phi \rightarrow \pi -\phi$ and
therefore $\sum_{i=1}^N \phi_i= (N-2)\pi \rightarrow \sum_{i=1}^N
\phi_i= 2\pi$.
}
Also when changing the $f$s while keeping fixed the $\epsilon$s the
shape may change as shown 
in fig.  (\ref{fig:change_f_N4M2}) for $N=4$ and $M_{ccw}=2$ and 
in fig.  (\ref{fig:change_f_N4M3}) for $N=4$ and $M_{ccw}=3$.
From now on we measure $M$ clockwise in not otherwise stated. 
Since the number
of reflex angles must be less or equal than $N-3$ we deduce that $
2\le M_{ccw} \le N-1$ or $1\le M_{cw} \le N-2$ 
and hence there are $N-2$ different sectors\footnote{
The symmetry 
$[X_{cl}(u,\bar u; \{1-\epsilon\}, \{ f^*\})]^*=X_{cl}(u,\bar u;
\{\epsilon\}, \{ f\})$  maps $M_{ccw}$ into $M_{cw}=N-M_{ccw}$ because it is
like the map $X\rightarrow X^*$ which reverses the order in which a
circuit is followed.
Hence  it does not map $M_{ccw}$ into $M_{ccw}'=N-M_{ccw}$.
.}

In the previous expressions (\ref{class-sols}) 
$\frac{\partial \omega_z}{\partial z}$
ensures the proper transformation under $SL(2,\R)$ and the product
$\prod_{j=2}^N (\omega_z-\omega_j)^{-(1-\epsilon_j)}$ and the
corresponding one for $\dz\cbX$ yields the proper monodromies around all
the point, $x_1$ included.
The extrema of the summations, i.e. the maximum allowed values of $n$
and $r$ are  chosen in order to have a finite
action and in particular their values are determined by the analysis of
the behavior of the solutions around $z=x_1$ and not around
$z=\infty$ as one would naively expect. This happens because the
solutions (\ref{class-sols})  behave  as
$O\left(\frac{1}{z^2}\right)$ at $z=\infty$ 
because of the factor $\frac{\partial  \omega_z}{\partial z}$. 

The powers of the products
$\prod_{j=2}^N(\omega_z-\omega_j)^{-(1-\epsilon_j)}$ 
and
$\prod_{j=2}^N(\omega_z-\omega_j)^{-\epsilon_j}$ 
are chosen in order to get a finite $X_{cl}(u,\bar u)$ 
at the singular points, explicitly using the definitions
(\ref{calX-calXbar}) and the expressions (\ref{class-sols}) 
we can write\footnote{
Because of way we have chosen the cuts we have 
$[(\omega_z-\omega_j)^\alpha]^*=(\omega_{\bar z}-\omega_j)^\alpha$
when $\omega_j$ is real.}
\begin{align}
X_{cl}(u,\bar u)
&=
f_1
&+
\sum_{n=0}^{N-M-2} a_n(\omega_j) 
\int_{x_1; z\in H}^u d z~
\frac{\partial \omega_z}{\partial z}
~\prod_{j=2}^N (\omega_z-\omega_j)^{-(1-\epsilon_j)}
\omega_z^n
\nonumber\\
&
&+
\sum_{r=0}^{M-2} b_r(\omega_j) 
%\left[
\int_{\bar x_1; z\in H^-}^{\bar u} d  z~
\frac{\partial \omega_{ z}}{\partial  z}
~\prod_{j=2}^N (\omega_{ z}-\omega_j)^{-\epsilon_j}
\omega_{ z}^r
%\right]^*
%
%
\nonumber\\
&=
f_1
&+
\sum_{n=0}^{N-M-2} a_n(\omega_j) 
\int_{-\infty; \omega\in H}^{\omega_u} d \omega~
~\prod_{j=2}^N (\omega-\omega_j)^{-(1-\epsilon_j)}
\omega^n
\nonumber\\
&
&+
\sum_{r=0}^{M-2} b_r(\omega_j) 
\left[
\int_{-\infty; \omega\in H}^{\omega_u} d \omega~
~\prod_{j=2}^N (\omega-\omega_j)^{-\epsilon_j}
\omega^r
\right]^*
\label{Xcl}
\end{align}
where in the last step we have used the explicit definition  of the power
to connect the integral performed in lower half plane with that
performed in the upper half plane.
In a similar way we can write
\begin{align}
\bar X_{cl}(u,\bar u)
&=
f_1^*
&+
e^{-i 2\pi \alpha_2}
\sum_{n=0}^{N-M-2} a_n(\omega_j) 
\left[
\int_{-\infty; \omega\in H}^{\omega_u} d \omega~
~\prod_{j=2}^N (\omega-\omega_j)^{-(1-\epsilon_j)}
\omega^n
\right]^*
\nonumber\\
&
&+
e^{-i 2\pi \alpha_2}
\sum_{r=0}^{M-2} b_r(\omega_j) 
\int_{-\infty; \omega\in H}^{\omega_u} d \omega~
~\prod_{j=2}^N (\omega-\omega_j)^{-\epsilon_j}
\omega^r
\label{barXcl}
\end{align}
where the coefficients are again $a$ and $b$ and not $a^*$ and $b^*$
as naively expected because we computed both
$X_{cl}$ and $\bar X_{cl}$ using the definitions of $\dz \cX$ and $\dz \cbX$ 
(\ref{calX-calXbar}) which mix both $\du X$ and $\dub \bX$ .
On the other side  $\bar X_{cl} = (X_{cl})^*$ hence there are  constraints on the
coefficients $a$ and $b$, i.e. $a^*=e^{-i 2\pi \alpha_2} a$ and
similarly for $b$ but these constraints are precisely the ones needed
to solve the equations (\ref{glob-bou-conds}) when one takes into
account the geometrical  requirements that $ f_{i+1}-f_{i}= e^{i \pi
\alpha_{i+1}} |f_{i+1}-f_{i}|$ as shown in fig. (\ref{fig:diff_f_angle}).

In order to determine the $N-M-1$ functions $a(\omega_j)$ ($j \ne 1,2,
N$) and the
$M-1$ $b(\omega_j)$ we
need simply to impose the $N-2$ geometrical constraints
\begin{equation}
X_{cl}(x_{i+1}, \bar x_{i+1})- X_{cl}(x_{i}, \bar x_{i})
= f_{i+1}-f_{i}
~~~~
i=2, \dots N-1
\label{glob-bou-conds}
\end{equation}
There is actually one more equation one can obviously impose, the one
with $i=1$ but it
turns out to be linearly dependent on the previous ones when the
geometrical constraints on $f$ and $\epsilon$ are imposed.
It is worth noticing that the previous constraints have an obvious
geometrical meaning differently from the the use of Pochammer path
used in  the literature.
The explicit solution of the previous constraints is given by solving
the linear system of $N-2$ equations\footnote{
The net effect of using the real valued integrals 
$I^{(N)}_{i,n}(\epsilon)$ is simply the phase 
$e^{+i \pi  \sum_{j=2}^i \epsilon_j}$ which multiply the real integral.
In particular for $I^{(N)}_{i,n}(1-\epsilon)$ we get
$(-1)^{i-1} e^{-i \pi  \sum_{j=2}^i \epsilon_j}$ and this explains the
alternating sign which appears weird at first sight.
}
\begin{align}
\sum_{n=0}^{N-M-2} (-)^{i-1} I^{(N)}_{i,n}(1-\epsilon_j) a_n
+
\sum_{r=0}^{M-2} I^{(N)}_{i,r}(\epsilon_j) b_r
=
e^{-i \pi \sum_{j=2}^i \epsilon_j} (f_i - f_{i+1})
~~~~
i=2,\dots N-1
\label{classical-solution-constraints}
\end{align}
where we have introduced the real valued integrals\footnote{
All these integrals can be expressed using $I^{(N)}_{i,0}$ since
$\omega_N=0$, explicitly 
$I^{(N)}_{i,n}(\alpha_j)=I^{(N)}_{i,0}(\alpha_j-\delta_{j,N} n)$ but we have
introduced this redundancy for notational simplicity.

Moreover we have used $\omega_{N+1}=\omega_1=-\infty$ because indexes
are defined $mod~N$.
}
\begin{align}
I^{(N)}_{i,n}(\alpha_j)
&=
\int_{\omega_{i+1}}^{\omega_i} d \omega~
~\prod_{j=2}^N |\omega-\omega_j|^{-\alpha_j}
\omega^n
\label{I-integral}
\end{align}
which are connected to type D Lauricella generalized hypergeometric
function when $i,i+1\ne 1$ by
\begin{align}
I^{(N)}_{i,0}(\alpha_j)
&=
\prod_{j\ne 1,i,i+1}^N |\omega_j -\omega_{i+1}|^{-\alpha_j}
~
|\omega_i -\omega_{i+1}|^{1-\alpha_{i+1}}
\nonumber\\
&~~~~
\cdot \int_0^1 d t ~t^{-\alpha_{i+1}} ~(1-t)^{-\alpha_{i}}
\prod_{j=2}^{i-1} 
\left(1-\frac{\omega_i-\omega_{i+1}}{\omega_j-\omega_{i+1}} t\right)^{-\alpha_{j}}
~\prod_{j=i+2}^{N} 
\left(1-\frac{\omega_i-\omega_{i+1}}{\omega_{i+1}-\omega_j}t \right)^{-\alpha_{j}}
\nonumber\\
&=
\prod_{j\ne 1,i,i+1}^N |\omega_j -\omega_{i+1}|^{-\alpha_j}
~
|\omega_i -\omega_{i+1}|^{1-\alpha_{i+1}-\alpha_{i}}
\frac{ 1
}{
B(1-\alpha_i,~ 1-\alpha_{i+1})
}
\nonumber\\
&~~~~
\cdot
F^{(N-3)}_D(1-\alpha_{i+1}; ~\{ \alpha_{j\ne 1,i,i+1} \};
~2-\alpha_i-\alpha_{i+1}; \xi_a)
\end{align}
where the parameters $\xi_a$ ($a=1,\dots N-3$) are given by
\begin{align}
\xi_a
&=
\left\{
\begin{array}{c c}
\frac{\omega_i-\omega_{i+1}}{\omega_{a+1}-\omega_{i+1}}
&
1\le a \le i-2
\\
\frac{\omega_i-\omega_{i+1}}{\omega_{a+3}-\omega_{i+1}}
& 
i-1\le a \le N-3
\end{array}
\right.
\end{align}
In particular 
for $N=3$ $F^{(0)}_D$ is Euler Beta function $B$ and
for $N=4$ $F^{(1)}_D$ is a plain hypergeometric $_2F_1$,
explicitly the previous expressions  become 
\begin{align}
I^{(3)}_{2,n}(1-\epsilon_j)
&=B(\epsilon_2, \epsilon_3+n)
\nonumber\\
I^{(4)}_{2,n}(1-\epsilon_j)
&=
\omega_3^{\epsilon_4-1+n}
~ (1-\omega_3)^{\epsilon_2+\epsilon_3-1}
~\frac{1}{B(\epsilon_3,\epsilon_2)}
~_2F_1(\epsilon_3; 1-n-\epsilon_4; \epsilon_2+\epsilon_3; 
\frac{\omega_3-1}{ \omega_3})
\nonumber\\
I^{(4)}_{3,n}(1-\epsilon_j)
&=
\omega_3^{\epsilon_3+\epsilon_4+n-1}
~\frac{1}{B(\epsilon_4+n, \epsilon_3)}
~_2F_1(1-\epsilon_2; n+\epsilon_4; \epsilon_3 +\epsilon_4 +n; 
\omega_3)
\label{I4}
\end{align}

We are now ready to compute the classical action for our solution.
Using the explicit expression for $\dz \cX$ and $\dzb \cbX$ we can write
\footnote{$e^{-i\pi\alpha_2}a_n$ is real as discussed before but it is
  by no means assured that it is positive.
}
\begin{align}
S_{cl}
=&
\frac{1}{8 \pi \alpha'}\Big[
\sum_{n,m=0}^{N-M-2}  (e^{-i\pi\alpha_2}a_n)~ (e^{-i\pi\alpha_2}a_m)
\int_{\C} d^2\omega~ 
\prod_{j=2}^N |\omega-\omega_j|^{-2(1-\epsilon_j)}
~\omega^n \bar \omega^m
\nonumber\\
&+
\sum_{r,s=0}^{M-2}  (e^{-i\pi\alpha_2}b_r) (e^{-i\pi\alpha_2}b_s)
\int_{\C} d^2\omega~ 
\prod_{j=2}^N |\omega-\omega_j|^{-2 \epsilon_j}
~\omega^r \bar \omega^s
\Big]
\end{align}
where an overall factor $\oh$ appears because we have extended the integration
domain from the upper half plane to the whole complex plane.
Notice that $e^{-i\pi\alpha_2} a, e^{-i\pi\alpha_2} b \in \R$ which
is however not enough to use their moduli $|a|$ and $|b|$ in the
previous expression.
As explained in appendix \ref{app:KLT} using the technique developed
in \cite{Kawai:1985xq} the previous integrals can be expressed as a product of
holomorphic and antiholomorphic contour integrals as
\begin{align}
&\int_{\C} d^2\omega~ 
\prod_{j=2}^N |\omega-\omega_j|^{-2 \epsilon_j}
~\omega^n \bar \omega^m
=
\nonumber\\
&~~=
\sum_{i=2}^{N-1}
\sum_{l=i+1}^{N}
\sin\left(\pi \sum_{j=i+1}^l \epsilon_j \right)
 I^{(N)}_{i,n}(\epsilon)
%\nonumber\\
%&\times
%\int_{\omega_{i+1}}^{\omega_i} d \xi
%\prod_{j=2}^N 
%|\xi -\omega_j|^{\alpha_j} (\xi -\omega_j)^{n_j}
%
%
%
%(-)^{\delta_{l,N}} 
I^{(N)}_{l,m}(\epsilon)
%
%\nonumber\\
%&\times
%\int_{\omega_{l+1}}^{\omega_l} d \eta
%\prod_{j=2}^N 
%|\omega_j-\eta|^{\alpha_j}
%(\eta-\omega_j)^{\bar n_j}
%
\end{align}

\subsection{The explicit $N=3$, $M_{cw}=1$ ($M_{ccw}=2$) case}
Let us start examining the $M_{cw}=1$ computation.
In this case we see immediately that $\dz \cbX$ is identically zero 
and that the only unknown is $a_0$ which is not a function but simply
a constant.
The eq.s (\ref{classical-solution-constraints}) reduce simply to
\begin{equation}
- a_0 e^{i\pi \epsilon_2} B(\epsilon_2, \epsilon_3)= f_2-f_3
\end{equation}
because $I^{(3)}_{2,n}(\alpha_j)= B(1-\alpha_2,1+n-\alpha_3)$ 
where $B(\epsilon_2, \epsilon_3)$ is Euler beta function.
The complete solution is then
\begin{equation}
X_{cl}^{(3,1_{cw})}(u,\bar u)
=
f_3
+ %
\frac{ e^{i\pi (1-\epsilon_2)} (f_2-f_3) 
}{
B(1, \epsilon_3) B(\epsilon_2, \epsilon_3)}
~_2F_1(\epsilon_3,~ 1-\epsilon_2;~ \epsilon_3+1;~ \omega_u)
~\omega_u^{\epsilon_3}
\end{equation}

Consider now the $M_{ccw}=2$ case where only $b_0$ is different from
zero and therefore $\dz \cX=0$. Proceeding as before we get
\begin{equation}
b_0 %e^{-i 2\pi \alpha_2} non dovrebbe esserci 
e^{i\pi \epsilon_2} B(1-\epsilon_2, 1-\epsilon_3)= f_2-f_3
\end{equation}
from which follows
\begin{equation}
X_{cl}^{(3,2_{ccw})}(u,\bar u)
=
f_3
+ %e^{-i\pi \epsilon_2}
\frac{ 
e^{-i\pi \epsilon_2}
(f_2-f_3) 
}{
B(1, 1-\epsilon_3) B(1-\epsilon_2, 1-\epsilon_3)}
~\left[_2F_1(1-\epsilon_3,~ \epsilon_2;~ 2-\epsilon_3;~ \omega_u)
~\omega_u^{1-\epsilon_3}\right]^*
\end{equation}
which explicitly shows the equivalence
$[X_{cl}^{(N,(N-M)_{cw})}(u,\bar u; \{1-\epsilon\}, \{ f^*\})]^*
=X_{cl}^{(N,M_{ccw})}(u,\bar u;\{\epsilon\}, \{ f\})$.

\subsection{The explicit $N=4$, $M_{cw}=1$ ($M_{ccw}=3$) case}
In this case we again immediately realize that $\dz \cbX$ is identically zero 
and that the only unknowns are the two functions 
$a_0(\omega_3)$ and $a_1(\omega_3)$.
The eq.s (\ref{classical-solution-constraints}) reduce simply to
\begin{align}
\sum_{n=0}^{1} (-)^{i-1} I^{(4)}_{i,n}(1-\epsilon_3) ~a_n(\omega_3)
=
e^{-i \pi \sum_{j=2}^i \epsilon_j} (f_i - f_{i+1})
~~~~
i=2,3
%\label{classical-solution-constraints}
\end{align}
with $I^{(4)}$s given explicitly in eq.s (\ref{I4})
The classical solution then  reads
\begin{align}
X_{cl}^{(4,1_{cw})}(u, \bar u)
&=
f_1
&+
\sum_{n=0}^{1} a_n(\omega_3) 
\int_{-\infty; \omega\in H}^{\omega_u} d \omega~
(\omega-1)^{\epsilon_2-1}
(\omega-\omega_3)^{\epsilon_3-1}
\omega^{n+\epsilon_4-1}
%~\prod_{j=2}^N (\omega-\omega_j)^{-(1-\epsilon_j)}
%\omega^n
\end{align}
The $X_{cl}^{(4,3_{ccw})}(u, \bar u)$ solution can then be obtained as
$
X_{cl}^{(4,3)}(u,\bar u;\{\epsilon\}, \{ f\})
=
[X_{cl}^{(4,1)}(u,\bar u; \{1-\epsilon\}, \{ f^*\})]^*
$.

\subsection{The explicit $N=4$, $M=2$ case}
In this case $M$ can be understood either as $M_{ccw}$ or as $M_{cw}$
which of the two can be only decided looking at the phases $\{\alpha_i\}$.
The unknowns are the two functions 
$a_0(\omega_3)$ and $b_0(\omega_3)$ and
eq.s (\ref{classical-solution-constraints}) reduce simply to
\begin{align}
(-)^{i-1} I^{(4)}_{i,0}(1-\epsilon_3) ~a_0(\omega_3)
+
I^{(0)}_{i,0}(\epsilon_3) ~b_r(\omega_3)
=
e^{-i \pi \sum_{j=2}^i \epsilon_j} (f_i - f_{i+1})
~~~~
i=2,3
%\label{classical-solution-constraintsN4M2}
\end{align}
hence the classical solution reads
\begin{align}
X_{cl}^{(4,2)}(u,\bar u)
&=
f_1
&+
a_0(\omega_3) 
\int_{-\infty; \omega\in H}^{\omega_u} d \omega~
(\omega-1)^{\epsilon_2-1}
(\omega-\omega_3)^{\epsilon_3-1}
\omega^{\epsilon_4-1}
%~\prod_{j=2}^N (\omega_z-\omega_j)^{-(1-\epsilon_j)}\omega_z^n
\nonumber\\
&
&+
b_0(\omega_3) 
\left[
\int_{-\infty; \omega \in H}^{\omega_u} d \omega~
(\omega-1)^{-\epsilon_2}
(\omega-\omega_3)^{-\epsilon_3}
\omega^{-\epsilon_4}
%~\prod_{j=2}^N (\omega_{ z}-\omega_j)^{-\epsilon_j}\omega_{ z}^r
\right]^*
\end{align}

\subsection{Wrapping contributions}
The wrapping contributions have been studied in
\cite{Cremades:2003qj} for the N=3 case and in 
\cite{Abel:2003yx}  for
the case $M=N-2$ and there is not any difference among the different
$M$ values therefore the results obtained there are valid.
Let us anyhow quickly review them.
Given a minimal $N$-polygon in $T^2$ with vertexes $\{f_i\}$,
i.e. with all vertexes in the fundamental cell,  
we can consider non minimal polygons which wrap the $T^2$.
These can be easierly described
as polygons which have vertexes $\{ \tilde f_i \}$ 
in the covering $\R^2$ where $T^2 \equiv R^2 / \Lambda$ with
 the lattice defined as $\Lambda=\{ n_1 e_1+ n_2 e_2 | n_1,n_2\in \Z\}$.
These configurations give an additive contribution to the classical
path integral as
\begin{align}
\langle 
\sigma_{\epsilon_1,f_1}(x_1) \dots \sigma_{\epsilon_N,f_N}(x_N)
\rangle_{T^2}
=
\cN(x_i,\epsilon_i)
\sum _{\{\tilde  f_i\}}
 e^{-S_{E,cl}(x_i,\epsilon_i, \tilde f_i)}
\label{general_corr_from_quantum_and_classical_on_T2}
\end{align}
% $e^{-S_{cl}(f_i)}\rightarrow \sum _{\{\tilde  f_i\}}e^{-S_{cl}(\tilde f_i)}$.

In order to determine the possible vertexes  $\{ \tilde f_i \}$
without redundancy it is necessary to keep a vertex fixed and then
expand the polygon. For definiteness we keep fixed the vertex $\tilde f_1=f_1$ 
which lies at the intersection between $D_N$ and $D_1$.
We then move the next vertex $f_2$ along the $D_1$ brane. 
Explicitly we write $\tilde f_2= \tilde f_1+ (f_2-f_1) + n_1 t_1= 
f_2+ n_1 t_1$ with $n_1\in \Z$ and
$t_1$  the shortest tangent vector to $D_1$ which is in $\Lambda$.
We can now continue for all the other vertexes for which we have
$\tilde f_i = \tilde f_{i-1} +( f_{i}-f_{i-1})+ n_{i-1} t_{i-1}
=f_i + \sum_{k=1}^{i-1} n_k t_k$.
For consistency we need requiring $\tilde f_{N+1} \equiv \tilde f_1 =
f_1$,
therefore the possible wrapped polygons are obtained from the solution
of the Diophantine equation
\begin{align}
\sum_{i=1}^N n_i t_i =0
\end{align}
which cannot be solved in general terms but only on a case by case
basis as discussed in \cite{Abel:2003yx}.
   
%%%%%%%%%%%%%%%%%%%%%%%%%%%%%%%%%%%%%%%%%%%%%%%%%%%%%%%%%%%%%%%%%%%%%%
%%%%%%%%%%%%%%%%%%%%%%%%%%%%%%%%%%%%%%%%%%%%%%%%%%%%%%%%%%%%%%%%%%%%%%
%%%%%%%%%%%%%%%%%%%%%%%%%%%%%%%%%%%%%%%%%%%%%%%%%%%%%%%%%%%%%%%%%%%%%%

\section{Green functions for $N\ge 3$}
\label{sect:Green_functions}
Having determined the classical solution we now compute the Green
functions in presence of twist fields both as an intermediate step
toward the computation of the quantum part of correlators 
and as a key ingredient to the computation of excited twist fields
correlators.

Following partially the literature we define the following quantities
for the quantum fluctuations 
which are connected with the derivatives of the Green functions as
\begin{align}
g_{(N,M)}(z, w; \{x_i\} )
&=
\frac{
\langle 
\dz \cX_q(z) \dw \cbX_q(w) \sigma_{\epsilon_1,f}(x_1) \dots \sigma_{\epsilon_N,f}(x_N)
\rangle
}{
\langle 
\sigma_{\epsilon_1,f}(x_1) \dots \sigma_{\epsilon_N,f}(x_N)
\rangle
}
\nonumber\\
h_{(N,M)}(z, w; \{x_i\} )
&=
\frac{
\langle 
\dz \cbX_q(z) \dw \cbX_q(w) \sigma_{\epsilon_1,f}(x_1) \dots \sigma_{\epsilon_N,f}(x_N)
\rangle
}{
\langle 
\sigma_{\epsilon_1,f}(x_1) \dots \sigma_{\epsilon_N,f}(x_N)
\rangle
}
\nonumber\\
l_{(N,M)}(z, w; \{x_i\} )
&=
\frac{
\langle 
\dz \cX_q(z) \dw \cX_q(w) \sigma_{\epsilon_1,f}(x_1) \dots \sigma_{\epsilon_N,f}(x_N)
\rangle
}{
\langle 
\sigma_{\epsilon_1,f}(x_1) \dots \sigma_{\epsilon_N,f}(x_N)
\rangle
}
\end{align}
We do not need to
consider functions involving antiholomorphic quantities
because $\dzb \cX$ and $\dzb \cbX$ are related to 
$\dz \cX$ and $\dz \cbX$ as in eq.s (\ref{dcX-dbcX}).

Quantum fluctuations are required to satisfy the boundary conditions
\begin{align}
Re( e^{-i \pi \alpha_{i}} \partial_y X_{q} |_{y=0} ) 
=
Im( e^{-i \pi \alpha_{i}} X_{q} |_{y=0}) 
=0
~~~~
x_{i+1}<x<x_i
.
\end{align}
These conditions that can reformulated as a set of local constraints
\begin{align}
\partial\cX_q(x_i+ e^{i 2\pi} \delta )
&= 
e^{i 2 \pi \epsilon_i}\partial\cX_q(x_i +\delta),
~~~~
%\nonumber\\
\partial\cbX_q(x_i+ e^{i 2\pi} \delta ) 
= 
e^{-i 2 \pi \epsilon_i} \partial\cbX_q(x_i +\delta)
\label{Gr-local-const}
\end{align}
and as a set of global constraints
\begin{align}
X_q(x_i,\bar x_i)
&=
X_q(x_{i+1},\bar x_{i+1}),
~~~~
\bar X_q(x_i,\bar x_i)
=
\bar X_q(x_{i+1},\bar x_{i+1}),
\label{Gr-global-const}
\end{align}
\COMMENTOOK{
Why not that the transverse fluctuations wrt the branes are zero?
I.e. from eq. (\ref{local-boundary-condition-s=0})
}

In the spirit of what done in the previous section we use a $SL(2,\R)$
invariant formulation and we write
\begin{align}
g_{(N,M)}(z, w; \{x_i\} )
&=
\frac{1}{(z-w)^2}
\prod_{j\ne 1}
\frac{(\omega_z-\omega_j)^{\epsilon_j-1}}{(\omega_w-\omega_j)^{\epsilon_j}}
\sum_{n=0}^{N-M}\sum_{s=0}^{M} a_{n s}(\omega_j) \omega_z^n \omega_w^s
\nonumber\\
h_{(N,M)}(z, w; \{x_i\} )
&=
e^{-i 2\pi \alpha_2}
\frac{\partial \omega_z}{\partial z}
\frac{\partial \omega_w}{\partial w}
\sum_{r,s=0}^{M-2} b_{r s}(\omega_j) 
~\dz_\omega \cbX^{(r)}(\omega_z)
~\dz_\omega \cbX^{(s)}(\omega_w)
\nonumber\\
l_{(N,M)}(z, w; \{x_i\} )
&=
e^{i 2\pi \alpha_2}
\frac{\partial \omega_z}{\partial z}
\frac{\partial \omega_w}{\partial w}
\sum_{n,m=0}^{N-M-2} c_{n m}(\omega_j) 
~\dz_\omega \cX^{(n)}(\omega_z)
~\dz_\omega \cX^{(m)}(\omega_w)
\end{align}
%%%%%%%%%%%%%%%%%%%%%%%%%%%%% COMMENTO
\COMMENTO{
\begin{align}
g_{(N,M)}(z, w; \{x_i\} )
&=
\frac{1}{(z-w)^2}
\prod_{j\ne 1}
\frac{(\omega_z-\omega_j)^{\epsilon_j-1}}{(\omega_w-\omega_j)^{\epsilon_j}}
\sum_{r=0}^{N-M}\sum_{s=0}^{M} c_{r s}(\omega_j) \omega_z^r \omega_w^s
\nonumber\\
h_{(N,M)}(z, w; \{x_i\} )
&=
\frac{\partial \omega_z}{\partial z}
\frac{\partial \omega_w}{\partial w}
\prod_{j\ne 1}
[(\omega_z-\omega_j)^{-\epsilon_j} (\omega_w-\omega_j)^{-\epsilon_j}]
\sum_{n,m=0}^{M-2} a_{n m}(\omega_j) \omega_z^n \omega_w^m
\nonumber\\
l_{(N,M)}(z, w; \{x_i\} )
&=
\frac{\partial \omega_z}{\partial z}
\frac{\partial \omega_w}{\partial w}
\prod_{j\ne 1}
[(\omega_z-\omega_j)^{-1+\epsilon_j} (\omega_w-\omega_j)^{-1+\epsilon_j}]
\sum_{n,m=0}^{N-M-2} b_{n m}(\omega_j) \omega_z^n \omega_w^m
\end{align}
} %fine commento
%%%%%%%%%%%%%%%%%%%%%%%%%%%%% COMMENTO
where $a_{n s}(\omega_j)$, $ b_{r s}(\omega_j)$ and $ c_{n  m}(\omega_j)$ 
are unknown functions of the anharmonic ratios
$\omega_{j\ne 1,2,N}$.

Let us rapidly review the ingredients of the previous construction. 
The factors $\frac{1}{(z-w)^2}$, $\frac{\partial \omega_z}{\partial z}
\frac{\partial \omega_w}{\partial w}$ and $\frac{\partial \omega_z}{\partial z}
\frac{\partial \omega_w}{\partial w}$ are there to ensure the proper
$SL(2,\R)$ transformations.
The powers of the singular parts have been chosen in order to 
%be able to choose the previous unknown functions in order to 
reproduce the singularities of OPEs
\begin{align}
\dz \cX(z) \dw \cbX(w) &\sim \frac{1}{(z-w)^2} + O(1)
\label{zw-sing-OPEs}
\\
\dz \cX(z) \sigma_{\epsilon,f}(x) &\sim (z-x)^{\epsilon-1} (\dz \cX \sigma_{\epsilon,f})(x)
\nonumber\\
\dz \cbX(z) \sigma_{\epsilon,f}(x) &\sim (z-x)^{-\epsilon} (\dz \cbX \sigma_{\epsilon,f})(x)
\label{sing-OPEs}
\end{align}
where $(\dz \cX \sigma_{\epsilon,f})(x)$ and $(\dz \cbX
\sigma_{\epsilon,f})(x)$ are excited twists.
In particular eq.s (\ref{Gr-local-const}) are the same of
eq.s (\ref{sing-OPEs}) as it should be since twist operators have been
introduced to generate (\ref{Gr-local-const}) constraints.

The upper bounds of the summation ranges are fixed by request that singularities
$z\rightarrow x_1$ and $w\rightarrow x_1$ are not worse than those in
eq.s (\ref{sing-OPEs}) while the lower bound is fixed from the
$z\rightarrow x_N$ and $w\rightarrow x_N$ limits.

There is another consistency condition: when
$x_i \rightarrow x_j$ 
we must obtain the corresponding Green function with $N\rightarrow N-1$.
It is worth stressing that at first sight there is a further constraint.
Usually the OPE between two twists is written as
\begin{align}
\sigma_{\epsilon_i, f}(x_i) ~\sigma_{\epsilon_j, f}(x_j)\sim
\left\{
\begin{array}{c c}
(x_i-x_j)^{- \epsilon_i \epsilon_j} ~\cM(\epsilon_i,\epsilon_j) ~\sigma_{\epsilon_i+\epsilon_j, f}(x_j)
&
\epsilon_i+\epsilon_j<1
\\
(x_i-x_j)^{- (1-\epsilon_i)(1- \epsilon_j)} ~\cN(\epsilon_i,\epsilon_j) ~\sigma_{\epsilon_i+\epsilon_j-1, f}
&
\epsilon_i+\epsilon_j>1
\end{array}
\right.
%+\mbox{descendants}
\label{sigma-sigma-OPE0}
\end{align}
with $\cM(\epsilon_i,\epsilon_j)=\cN(\epsilon_i,\epsilon_j)=1$.
We will discuss that it is not possible to set both $\cM$ and $\cN$ to one
in section \ref{sub:N-1-from-N}.
Now we would however comment on the fact that the previous expression is
written without
higher terms leading to the wrong impression that all the omitted 
terms are descendants.
If it were true that the OPE (\ref{sigma-sigma-OPE0}) has no other
primaries in the rhs 
this would imply that the derivatives of 
Green functions are analytic functions of
the variables $x_i$ too since the overall singularity in $x_i - x_j$ 
due to the power factor would cancel between the numerator and the denominator.
This  is  not true as the explicit computations show but in the
$M_{cw}=1$, $M_{ccw}=N-1$
case and the reason is that the previous OPE involves actually an infinite
number of primary fields with powers of OPE coefficients which do not
differ by integers, explicitly for $\epsilon_i+\epsilon_j<1$
\begin{align}
\sigma_{\epsilon_i, f}(x_i) \sigma_{\epsilon_j, f}(x_j)
\sim
&
(x_i-x_j)^{\epsilon_i \epsilon_j}~
\cM(\epsilon_i, \epsilon_j)~ 
\sigma_{\epsilon_i+\epsilon_j, f}(x_j)
\nonumber\\
&+ 
\sum_{k=1} c_k ~
(x_i-x_j)^{\epsilon_i \epsilon_j+ k(\epsilon_i+\epsilon_j) } 
[(\partial X)^k   \sigma_{\epsilon_i+\epsilon_j, f}](x_j)
+\dots
\end{align}
where $c_k$ are certain numbers and $\dots$ stands for other primaries
and descendants.
These primary fields have a simple interpretation as the states
associated to the Hilbert space of twisted string since all of them
have conformal dimensions which differ by multiples of $\pm \epsilon$.

To continue and write in a more compact way the following expressions we define
\begin{align}
P(\omega_z, \omega_w)
&=
\prod_{j=2}^N
\frac{(\omega_z-\omega_j)^{\epsilon_j-1}}{(\omega_w-\omega_j)^{\epsilon_j}}
\nonumber\\
S(\omega_z, \omega_w)
&=
\sum_{n=0}^{N-M}\sum_{s=0}^{M} a_{n s}(\omega_j) \omega_z^n \omega_w^s
\label{def-S-P}
\end{align}
When we impose the constraint from the $z\rightarrow w$ limit 
given in eq. (\ref{zw-sing-OPEs}) we get 
\begin{align}
S(\omega_w, \omega_w)
&= 
\sum_{n=0}^{N-M}\sum_{s=0}^{M} a_{n s}(\omega_j) \omega_w^{n+s}
=
\prod_{j=2}^N (\omega_w - \omega_j)
\nonumber\\
\frac{\partial S}{\partial \omega_z}\Big|_{\omega_z=\omega_w}
&=
\sum_{n=0}^{N-M}\sum_{s=0}^{M} n a_{n s}(\omega_j) \omega_w^{n+s-1}
=
\sum_{j=2}^N \frac{1-\epsilon_j}{\omega_w-\omega_j}
\cdot
\prod_{l=2}^N (\omega_w - \omega_l)
\label{S-dS-const}
\end{align}
or the equivalent equations with $w\rightarrow z$ and $\epsilon
\rightarrow 1-\epsilon$.
These are $(N+1)+N$ equations for $a_{n s}$ but only 
$2 N$ are independent since both imply that $a_{N-M,~M}=0$.
Generically, i.e. for $M_{cw} \ne 1$ 
these equations are not sufficient to fix the $(N-M+1)(M+1)$
unknowns $a_{n s}$ and must be supplemented by the constraints which follow
from eq.s (\ref{Gr-global-const}).
These further constraints allow also to fix the
remaining $(M-1)^2+(N-M-1)^2$ unknowns functions $b_{r s}$ and $c_{n m}$.
For example from the first equation in (\ref{Gr-global-const}) we get
\begin{align}
X_q(x_i,\bar x_i)-X_q(x_{i+1},\bar x_{i+1})
&=
\int_{x_{i+1}}^{x_i} d X_q
=
\int_{x_{i+1}}^{x_i} dx [ \du X_q(x+i 0^+)+ \dub X_q(x-i 0^+)]
\nonumber\\
&=
\int_{x_{i+1}}^{x_i} dx [ \dz \cX_q(x+i 0^+)+  e^{i 2\pi\alpha_2} \dzb
  \cbX_q(x-i 0^+)]
=0
\end{align}
which implies the constraints\footnote{
It is worth noticing that the segment $[x_{i+1},x_i]$ is followed for
one addend above and for the other below the cut (it works also the
other way round w.r.t. the main text). This ensures that
both addends have the same phase modulus $\pi$.
Consistency among possible formulations of the constraints is due to 
$[g(z,w)]^*=g(\bar z, \bar w)$,
$[h(z,w)]^*= e^{i 4\pi  \alpha_2} h(\bar z, \bar w)$ and
$[l(z,w)]^*= e^{-i 4\pi  \alpha_2} l(\bar z, \bar w)$.
}
\begin{align}
&\int_{x_{i+1}}^{x_{i}} d x~ g_{(N,M)}(x+i 0^+,w)+ e^{i 2\pi \alpha_2} h_{(N,M)}(x-i 0^+,w)=0
\nonumber\\
&\int_{x_{i+1}}^{x_{i}} d x~   l_{(N,M)}(z,x-i 0^+) 
+ e^{i 2\pi  \alpha_2} g_{(N,M)}(z,x+i 0^+)=0
\label{glob-consts-gh-gl}
\end{align}
These can be explicitly written as
\begin{align}
\sum_{s=0}^M \omega_w^s
& 
\sum_{n=0}^{N-M} a_{n s}(\omega_j)
\int_{\omega_{i+1}; \omega\in H}^{\omega_{i}} \frac{ d \omega}{(\omega-\omega_w)^2}
\prod_{j=2}^N (\omega-\omega_j)^{\epsilon_j-1}
~\omega^n
\nonumber\\
&+
\sum_{s=0}^{M-2} \omega_w^s 
\sum_{r=0}^{M-2} b_{r s}(\omega_j)
\int_{\omega_{i+1}; \omega\in H^-}^{\omega_{i}} { d \omega}
\prod_{j=2}^N (\omega-\omega_j)^{-\epsilon_j}
~\omega^r
=0
\label{Green-global-constraints0-gh}
\\
\sum_{n=0}^{N-M} \omega_z^n
& 
\sum_{s=0}^{M} a_{n s}(\omega_j)
\int_{\omega_{i+1}; \omega\in H^-}^{\omega_{i}} \frac{ d \omega}{(\omega_z-\omega)^2}
\prod_{j=2}^N (\omega-\omega_j)^{-\epsilon_j}
~\omega^s
\nonumber\\
&+
\sum_{n=0}^{N-M-2} \omega_z^n 
\sum_{m=0}^{N-M-2} c_{n m}(\omega_j)
\int_{\omega_{i+1}; \omega\in H}^{\omega_{i}} { d \omega}
\prod_{j=2}^N (\omega-\omega_j)^{\epsilon_j-1}
~\omega^m
=0
\label{Green-global-constraints0-gl}
\end{align}
As in the case for the classical solution only
$N-2$ of the previous intervals give independent constraints, let us
say $i=2,\dots N-1$. 
All these constraints are then sufficient to fix completely and
uniquely all the coefficients.
These constraints are actually much more than needed since they equate
a polynomial in $\omega_w$ or in $\omega_z$ to an analytic function.
If we expand around $\omega_w=\omega_z=\infty$ 
and consider only the polynomial part we have enough
equations to fix all the unknowns since to the previous $2 N$ constraints
in eq.s (\ref{S-dS-const}) we add  
$M-1$ equations in $\omega_w$  times $N-2$ intervals and 
$N-M-1$ equations for $\omega_z$ times $N-2$ intervals.
%\COMMENTOO{
Actually the previous equations are already overdetermined in  $a$
since the $2 N$ eq.s (\ref{S-dS-const}) and the $(N-2)(M-1)$ ones in
$\omega_w$ are sufficient for fix both $a$ and $b$ and similarly for
the ones in $\omega_z$
%Pensare al caso $N_{cw}=1$
%}
therefore for consistency  we suppose 
that this overdetermined system is consistent as well as 
all the remaining equations obtained from the polar part in
$\omega_w$ and $\omega_z$.
As far as the consistency of the functions $a$ determined in the two
ways we have checked it in particular limits in the explicit cases treated
afterward.
Moreover all constraints derived from the polar part 
are polynomials in the integrals $I^{(N)}$ (\ref{I-integral}) since the
functions $a$ and $b$ are solutions of a linear system whose
coefficients are precisely the $I^{(N)}$s.
Nevertheless it is easy to show that all constraints must be equivalent to
a relation with polynomial coefficients in
$\omega_{j\ne 1,2, N}$ and $I^{(N)}$ and at most linear in 
$\hat I^{(N+1)}$ (see eq. (\ref{Ihat-integral}))
with one of the parameters equal to $2$.
This can be seen as follows.
It is possible to split $g_{(N,M)}(z, w; \{x_i\} )$ in a singular part
and a regular one as 
\begin{align}
g_{(N,M)}(z, w; \{x_i\} )
&=
g_{s (N,M)}(z, w; \{x_i\} )
+g_{r (N,M)}(z, w; \{x_i\} )
\nonumber\\
g_{s (N,M)}(z, w; \{x_i\} )
&=
\frac{1}{(z-w)^2}
\prod_{j\ne 1}
\frac{(\omega_z-\omega_j)^{\epsilon_j-1}}{(\omega_w-\omega_j)^{\epsilon_j}}
\sum_{n=0}^{N-M}\sum_{s=0}^{M} a_{(0) n s}(\omega_j) \omega_z^n \omega_w^s
\nonumber\\
g_{r (N,M)}(z, w; \{x_i\} )
&=
\frac{\partial \omega_z}{\partial z}
\frac{\partial \omega_w}{\partial w}
\sum_{n=0}^{N-M-2}\sum_{r=0}^{M-2} \bar a_{n s}(\omega_j) 
~\dz_\omega \cX^{(n)}(\omega_z)
~\dz_\omega \cbX^{(r)}(\omega_w)
\label{g-gs-gr}
\end{align}
This splitting is completely arbitrary and therefore it is not
uniquely defined  but it can be made unique imposing further
conditions such for example the request of setting to zero all the
$a_{n, s=k-n}$ but the two\footnote{
Eq.s (\ref{S-dS-const}) divide naturally the set of unknowns 
$\{a_{n  s}\}$ into subsets $\{a_{n  s}\}_{s+n=k}$ and for each of
these subsets there are two linear equations.
} with lowest $n$ as for example in
eq. (\ref{g42-singular-lowest-n}) for the $(N=4, M=2)$ case
or 
%\COMMENTOO{dire meglio?}
the request that the singular part $g_{s  (N,M)}$ 
goes into $g_{s  (N-1,M')}$ when $x_i \rightarrow x_j$
as shown in appendix \ref{app:g_sing}  and explicitly in
eq. (\ref{g42-singular-n-into-n-1}) for $(N=4, M=2)$ case.
 \COMMENTO{PIU' LIMITE -DA VERIFICARE- altrimenti ci sono delle
  indeterminazioni quindi e' dar dire bene}  
Once fixed by a ``gauge choice'' the singular part 
the regular one is fixed by the global boundary conditions.

Actually if we choose to split $g$ into a regular part and a singular
part (which is fixed not uniquely by the OPEs) as in eq.s
(\ref{g-gs-gr}) the equation
(\ref{Green-global-constraints0-gh}) can be written as
\begin{align}
\sum_{s=0}^M \omega_w^s
& 
\sum_{n=0}^{N-M} a_{(0) n s}(\omega_j)
\int_{\omega_{i+1}; \omega\in H}^{\omega_{i}} \frac{ d \omega}{(\omega-\omega_w)^2}
\prod_{j=2}^N (\omega-\omega_j)^{\epsilon_j-1}
~\omega^n
\nonumber\\
&+
\sum_{s=0}^{M-2} \omega_w^s
\sum_{n=0}^{N-M-2} \bar a_{ n s}(\omega_j)
\int_{\omega_{i+1}; \omega\in H}^{\omega_{i}} d \omega
%\prod_{j=2}^N (\omega-\omega_j)^{\epsilon_j-1}~\omega^n
\dz_\omega \cX^{(n)}(\omega)
\nonumber\\
&+
\sum_{s=0}^{M-2} \omega_w^s 
\sum_{r=0}^{M-2} b_{r s}(\omega_j)
\int_{\omega_{i+1}; \omega\in H^-}^{\omega_{i}} { d \omega}
%\prod_{j=2}^N (\omega-\omega_j)^{-\epsilon_j}~\omega^r
\dz_\omega \cbX^{(r)}(\omega)
=0
\label{Green-global-constraints-sign+reg-gh}
\end{align}
which reveals that 
the singular part $g_{s (N,M)}$  contributes with a term linear in 
$\hat I^{(N+1)}$  (with one parameter equal to $2$ because of the term
$(\omega-\omega_w)^{-2}$)
while the other terms have rational coefficient in
$\omega_{j\ne 1,2, N}$ and $I^{(N)}$ once we plug the solution for the
coefficients back.  
In a similar way we can write the equation corresponding to
(\ref{Green-global-constraints0-gl}) as
\begin{align}
\sum_{n=0}^{N-M} \omega_z^n
& 
\sum_{s=0}^{M} a_{(0) n s}(\omega_j)
\int_{\omega_{i+1}; \omega\in H}^{\omega_{i}} \frac{ d \omega}{(\omega_z-\omega)^2}
\prod_{j=2}^N (\omega-\omega_j)^{-\epsilon_j}
~\omega^n
\nonumber\\
&+
\sum_{n=0}^{N-M-2} \omega_z^n
\sum_{s=0}^{M-2} \bar a_{ n s}(\omega_j)
\int_{\omega_{i+1}; \omega\in H}^{\omega_{i}} d \omega
%\prod_{j=2}^N (\omega-\omega_j)^{\epsilon_j-1}~\omega^n
\dz_\omega \cbX^{(s)}(\omega)
\nonumber\\
&+
\sum_{m=0}^{N-M-2} \omega_z^m 
\sum_{n=0}^{N-M-2} b_{n m}(\omega_j)
\int_{\omega_{i+1}; \omega\in H^-}^{\omega_{i}} { d \omega}
%\prod_{j=2}^N (\omega-\omega_j)^{-\epsilon_j}~\omega^r
\dz_\omega \cX^{(n)}(\omega)
=0
\label{Green-global-constraints-sign+reg-gl}
\end{align}

Having determined the derivatives of the Green functions we can
reconstruct the actual Green functions as
\begin{align}
G^{ X \bX}_{(N,M)}(u, \ub; v, \vb; \{x_i\})
&=
\int_{x_i; u'\in H}^u d u' \int_{x_j; v'\in H}^v d v' ~g_{(N,M)}(u', v';  \{x_i\}))
\nonumber\\
&+
e^{-i 2\pi  \alpha_2}
\int_{x_i; u\in H}^u d u' \int_{x_j; \vb' \in H^-}^\vb d \vb' 
~l_{(N,M)}(u', \vb';  \{x_i\}))
\nonumber\\
&+
e^{i 2\pi  \alpha_2}
\int_{x_i; \ub'\in H^-}^\ub d \ub' \int_{x_j; v'\in H}^v d v' 
~h_{(N,M)}(\ub', v';  \{x\_i\}))
\nonumber\\
&+
\int_{x_i; \ub\in H^-}^\ub d \ub' \int_{x_j; \vb' \in H^-}^\vb d \vb' ~g_{(N,M)}(\vb', \ub';  \{x_i\}))
\end{align}
and
\begin{align}
G^{ X X}_{(N,M)}(u, \ub; v, \vb; \{x_i\})
&=
\int_{x_i; u'\in H}^u d u' \int_{x_j; v'\in H}^v d v' ~l_{(N,M)}(u', v';  \{x_i\}))
\nonumber\\
&+
e^{i 2\pi  \alpha_2}
\int_{x_i; u\in H}^u d u' \int_{x_j; \vb' \in H^-}^\vb d \vb' ~g_{(N,M)}(u', \vb';  \{x_i\}))
\nonumber\\
&+
e^{i 2\pi  \alpha_2}
\int_{x_i; \ub'\in H^-}^\ub d \ub' \int_{x_j; v'\in H}^v d v' ~g_{(N,M)}(v', \ub';  \{x_i\}))
\nonumber\\
&+
e^{i 4\pi  \alpha_2}
\int_{x_i; \ub\in H^-}^\ub d \ub' \int_{x_j; \vb' \in H^-}^\vb d \vb' ~h_{(N,M)}(\ub', \vb';  \{x_i\}))
\end{align}
and
\begin{align}
G^{ \bX \bX}_{(N,M)}(u, \ub; v, \vb; \{x_i\})
&=
\int_{x_i; u'\in H}^u d u' \int_{x_j; v'\in H}^v d v' ~h_{(N,M)}(u', v';  \{x_i\}))
\nonumber\\
&+
e^{-i 2\pi  \alpha_2}
\int_{x_i; u\in H}^u d u' \int_{x_j; \vb' \in H^-}^\vb d \vb'
~g_{(N,M)}(\vb', u';  \{x_i\}))
\nonumber\\
&+
e^{-i 2\pi  \alpha_2}
\int_{x_i; \ub'\in H^-}^\ub d \ub' \int_{x_j; v'\in H}^v d v'
~g_{(N,M)}(\ub', v';  \{x_i\}))
\nonumber\\
&+
e^{-i 4\pi  \alpha_2}
\int_{x_i; \ub\in H^-}^\ub d \ub' \int_{x_j; \vb' \in H^-}^\vb d \vb' 
~l_{(N,M)}(\ub', \vb';  \{x_i\}))
\end{align}
where the arbitrariness of the lower integration limit is due to the
constraints (\ref{glob-consts-gh-gl}) which allow to change
$x_i\rightarrow x_k$.

\subsection{The explicit $N=3$, $M_{cw}=1$ case}
In this case $g_{(3,1)}$ is completely fixed by the local constraints 
to be
\begin{align}
g_{(3,1)}(z,w;\{x_i\})
&=
\frac{1}{(z-w)^2}
\frac{ (\omega_z-1)^{\epsilon_2-1} }{ (\omega_w-1)^{\epsilon_2} }
\frac{ \omega_z^{\epsilon_3-1} }{ \omega_w^{\epsilon_3} }
\nonumber\\
&
\Big[
(1-\epsilon_2-\epsilon_3) \omega_z^2 
+(\epsilon_2+\epsilon_3) \omega_z \omega_w
-(1-\epsilon_3) \omega_z -\epsilon_3 \omega_w
\Big]
\end{align}
while $h_{(3,1)}=0$ since $M=1$.
We get therefore a constraint from
eq. (\ref{Green-global-constraints0-gh}) or the equivalent form
(\ref{Green-global-constraints-sign+reg-gh}) which read
\begin{align}
\omega_w 
&
\int_0^1 d\omega~
\frac{1}{(\omega-\omega_w)^2}
(\omega-1)^{\epsilon_2-1}
~ \omega^{\epsilon_3-1}
[ (\epsilon_2+\epsilon_3) \omega - \epsilon_3]
\nonumber\\
&+
\int_0^1 d\omega~
\frac{1}{(\omega-\omega_w)^2}
(\omega-1)^{\epsilon_2-1}
~ \omega^{\epsilon_3-1}
[ (1-\epsilon_2-\epsilon_3) \omega^2 -(1- \epsilon_3) \omega]
=0
\end{align}
which can be read either as a constraint on the hypergeometric
functions
\begin{align}
&
(1-\epsilon_2-\epsilon_3) 
  ~B(2+\epsilon_3, \epsilon_2) 
  ~_2 F_1( 2, 2+\epsilon_3; 2+\epsilon_2+\epsilon_3;  \frac{1}{\omega_w})
\nonumber\\
&
-
(1-\epsilon_3) 
  ~B(1+\epsilon_3, \epsilon_2) 
  ~_2 F_1( 2, 1+\epsilon_3; 1+\epsilon_2+\epsilon_3;  \frac{1}{\omega_w})
\nonumber\\
&
+
\left[
(\epsilon_2+\epsilon_3) 
  ~B(1+\epsilon_3, \epsilon_2) 
  ~_2 F_1( 2, 1+\epsilon_3; 1+\epsilon_2+\epsilon_3;  \frac{1}{\omega_w})
-
\epsilon_3
  ~B(\epsilon_3, \epsilon_2) 
  ~_2 F_1( 2, 1+\epsilon_3; \epsilon_2+\epsilon_3;  \frac{1}{\omega_w})
\right] \omega_w
=0
\label{const-BF}
\end{align}
or as  
infinite constraints on the coefficients of the $\omega_w$ expansion
which relate different Beta functions. 
We are now left to determine $l_{(3,1)}$ from
eq. (\ref{Green-global-constraints0-gl}), explicitly
\begin{align}
&
\omega_z^2
~
\int_0^1 d\omega~
\frac{1}{(\omega_z-\omega)^2}
(\omega-1)^{-\epsilon_2}
~ \omega^{-\epsilon_3}
(1-\epsilon_2-\epsilon_3) 
\nonumber\\
+
&\omega_z
~
\int_0^1 d\omega~
\frac{1}{(\omega_z-\omega)^2}
(\omega-1)^{-\epsilon_2}
~ \omega^{-\epsilon_3}
[ (1-\epsilon_2-\epsilon_3) \omega^2 -(1- \epsilon_3) \omega]
\nonumber\\
+
&
\int_0^1 d\omega~
\frac{1}{(\omega_z-\omega)^2}
(\omega-1)^{-\epsilon_2}
~ \omega^{-\epsilon_3}
(-\epsilon_3 \omega)
\nonumber\\
+
&
c_{0 0}
~
\int_0^1 d\omega~
(\omega-1)^{\epsilon_2-1}
~ \omega^{\epsilon_3-1}
=
0
\label{const-l31}
\end{align}
or
\begin{align}
&
\omega_z^2 ~(1-\epsilon_2-\epsilon_3) ~\hat I^{(4)}_{2,0}(\epsilon_j; 2)
\nonumber\\
+
&\omega_z~
[ (1-\epsilon_2-\epsilon_3) ~\hat I^{(4)}_{2,2}(\epsilon_j; 2) 
-(1- \epsilon_3) ~\hat I^{(4)}_{2,1}(\epsilon_j; 2) ]
\nonumber\\
-
&
\epsilon_3 ~\hat I^{(4)}_{2,1}(\epsilon_j; 2)
%
%\nonumber\\
+
c_{0 0}
~
I^{(3)}_{2,0}(\epsilon_j)
%B(\epsilon_2, \epsilon_3)
=
0
\label{const-l31-I}
\end{align}
where we have introduced the function 
%%%%%$\hat I^{(5)}_{i,0}(1-\epsilon_j ; \beta)$ 
\begin{align}
\hat I^{(N)}_{i,n}(\alpha_j; \beta)
&=
\int_{\omega_{i+1}}^{\omega_i} d \omega~
~\prod_{j=2}^N |\omega-\omega_j|^{-\alpha_j}
~(\omega-\omega_w)^{-\beta}
\omega^n
 \label{Ihat-integral}
\end{align}
which is a slight modification of our previous definition
(\ref{I-integral}) and is still 
connected to the Lauricella functions $F_D^{(n)}$.
In the $\omega_z \rightarrow \infty$ limit we can determine the
unique unknown coefficient and hence the $l_{(3,1)}$ normalization to be  
\begin{align}
c_{0 0}
=
- (1-\epsilon_2 -\epsilon_3)  
\frac{ B(1-\epsilon_2, 1-\epsilon_3) }{ B(\epsilon_2, \epsilon_3)}
\end{align}
We get also infinite constraints from the subleading orders in
$\omega_z$ or plugging the previous value for $c_{0 0}$ 
back into eq. (\ref{const-l31})
an equation of the form $\sum B ~_2 F_1=0$ as eq. (\ref{const-BF}).

\subsection{The explicit $N=4$, $M=1$ case}
Again as the case before $g_{(4,1)}$ is completely fixed by the local
constraints only to be
\begin{align}
g_{(4,1)}(z,w;\{x_i\})
&=
\frac{1}{(z-w)^2}
\frac{ (\omega_z-1)^{\epsilon_2-1} }{ (\omega_w-1)^{\epsilon_2} }
\frac{ (\omega_z-\omega_3)^{\epsilon_3-1} }{ (\omega_w-\omega_3)^{\epsilon_3} }
\frac{ \omega_z^{\epsilon_4-1} }{ \omega_w^{\epsilon_4} }
\nonumber\\
&
\Big[
\epsilon_1 \omega_z^3 
+ 
  (1-\epsilon_1) \omega_z^2 \omega_w
-
 [ (1-\epsilon_3-\epsilon_4) +  (1-\epsilon_2-\epsilon_4) \omega_3  
 ]\omega_z^2 
\nonumber\\
&
-
 [(\epsilon_3+\epsilon_4)+ (\epsilon_2+\epsilon_4) \omega_3 
 ]\omega_z \omega_w
+
(1-\epsilon_4) \omega_3 \omega_z 
+\epsilon_4 \omega_3 \omega_w
\Big]
\end{align}
and $h_{(4,1)}=0$ since $M=1$.
As in the $(3,1)$ case
from eq. (\ref{Green-global-constraints0-gh}) or the equivalent form
(\ref{Green-global-constraints-sign+reg-gh}) we get  the constraints
\begin{align}
&
-
\omega_w^2
 [ (1-\epsilon_3-\epsilon_4) +  (1-\epsilon_2-\epsilon_4) \omega_3  
 ] ~\hat I^{(5)}_{i,0}(1-\epsilon_j ; 2)
\nonumber\\
&
+
\omega_w
\{ 
  (1-\epsilon_1) ~\hat I^{(5)}_{i,0}(1-\epsilon_j ; 2)
-
 [(\epsilon_3+\epsilon_4)+ (\epsilon_2+\epsilon_4) \omega_3 
 ] ~\hat I^{(5)}_{i,1}(1-\epsilon_j ; 2)
+
\epsilon_4 \omega_3 ~\hat I^{(5)}_{i,0}(1-\epsilon_j ; 2)
\}
\nonumber\\
&
+ \epsilon_1 ~\hat I^{(4)}_{i,3}(1-\epsilon_j)
+ (1-\epsilon_4) ~\omega_3 ~\hat I^{(4)}_{i,0}(1-\epsilon_j)
=
0
\end{align}
 In particular notice that
$\hat I^{(5)}_{i,n}\sim F_D^{(2)}$ is the Appell function.

We can proceed to determine the $l_{(4,1)}$ function. 
This amounts to fixing the four functions $c_{00}, c_{01}, c_{10},
c_{11}$ from eq. (\ref{Green-global-constraints0-gl}) which reads
\begin{align}
&
(-1)^{i+1}
\Big\{
\omega_z^3~
\epsilon_1  ~\hat I^{(5)}_{i,0}(\epsilon_j ; 2)
+ 
\omega_z^2~
  (1-\epsilon_1)  \hat I^{(5)}_{i,1}(\epsilon_j ; 2) 
-
\omega_z~
 [ (1-\epsilon_3-\epsilon_4) +  (1-\epsilon_2-\epsilon_4) \omega_3  
 ] \hat I^{(5)}_{i,1}(1-\epsilon_j ; 2) 
\nonumber\\
&
-
\omega_z~
 [(\epsilon_3+\epsilon_4)+ (\epsilon_2+\epsilon_4) \omega_3 
 ] \hat I^{(5)}_{i,1}(\epsilon_j ; 2) 
+
\omega_z~
(1-\epsilon_4) ~\omega_3 ~ \hat I^{(5)}_{i,0}(\epsilon_j ; 2)  
+\epsilon_4 \omega_3 \hat I^{(5)}_{i,1}(\epsilon_j ; 2)
\Big\}
\nonumber\\
&
+c_{00} ~I^{(4)}_{i,0}(1-\epsilon_j) 
+c_{01} ~I^{(4)}_{i,1}(1-\epsilon_j) 
+ \omega_z~c_{10} ~I^{(4)}_{i,0}(1-\epsilon_j) 
+ \omega_z~c_{11} ~I^{(4)}_{i,1}(1-\epsilon_j) 
=0
\label{const-l41}
\end{align}
for $i=2,3$. When we consider the $\omega_z \rightarrow \infty$ limit
we get two sets of equations, the one from the coefficient of $\omega_z$
\begin{align}
(-1)^{i+1}
\epsilon_1  ~I^{(4)}_{i,0}(\epsilon_j)
+ c_{10} ~I^{(4)}_{i,0}(1-\epsilon_j) 
+ c_{11} ~I^{(4)}_{i,1}(1-\epsilon_j) =0
\end{align}
and the other from the coefficient of $\omega_z^0$
\begin{align}
(-1)^{i+1}
(1+\epsilon_1)  ~\hat I^{(4)}_{i,1}(\epsilon_j)
+c_{00} ~I^{(4)}_{i,0}(1-\epsilon_j) 
+c_{01} ~I^{(4)}_{i,1}(1-\epsilon_j) 
=0
\end{align}
plus an infinite set of constraints from the coefficients of  the
polar expansion in $\omega_z$ or, equivalently
plugging  the previous value back into eq. (\ref{const-l41})
and equation of the form $ (_2 F_1)^2  F^{(2)}_D+ 
\sum(_2 F_1)^3=0$ analogously to eq. (\ref{const-BF}).

\subsection{The explicit $N=4$, $M=2$ case}
%\subsection{Discussion of the general case}
This is the first case where there are more unknown coefficients than equations from
the local constraints and therefore we must use the global constraints
to fix completely $g_{(4,2)}$ and determine both $h_{(4,2)}$ and
$l_{(4,2)}$ which are now both not vanishing.
We can nevertheless fix the singular part $g_{s (4,2)}$ by choosing
$a_{2 0}=0$ so we can get
\begin{align}
g_{s (4,2)}&=
\frac{1}{(z-w)^2}
\frac{ (\omega_z-1)^{\epsilon_2-1} }{ (\omega_w-1)^{\epsilon_2} }
\frac{ (\omega_z-\omega_3)^{\epsilon_3-1} }{ (\omega_w-\omega_3)^{\epsilon_3} }
\frac{ \omega_z^{\epsilon_4-1} }{ \omega_w^{\epsilon_4} }
\nonumber\\
&
\Big\{
\epsilon_1 \omega_z^2 \omega_w 
+ 
  (1-\epsilon_1) \omega_z \omega_w^2
-
 [ (2-\epsilon_3-\epsilon_4) +  (2-\epsilon_2-\epsilon_4) \omega_3  
 ]\omega_z \omega_w
\nonumber\\
&
+
 [(1-\epsilon_3-\epsilon_4)+ (1-\epsilon_2-\epsilon_4) \omega_3 
 ] \omega_w^2
+
(1-\epsilon_4) \omega_3 \omega_z 
+\epsilon_4 \omega_3 \omega_w
\Big\}
\label{g42-singular-lowest-n}
\end{align}
Using the global constraints for $g$ and $h$ as given in
eq. (\ref{Green-global-constraints-sign+reg-gh}) for $i=2,3$ it is then possible
to determine $\bar a_{0 0}$ (which corresponds to $a_{2 0}$  
after the split of $g_{(4,2)}$ into a regular and singular part)
and 
$b_{0 0}$, in particular taking the $\omega_w \rightarrow \infty$
limit we get
\begin{align}
\left\{
\begin{array}{c}
\bar a_{0 0} ~I^{(4)}_{2,0}(1-\epsilon)
-
b_{0 0} ~I^{(4)}_{2,0}(\epsilon)
=
-(1-\epsilon_1) ~I^{(4)}_{2,1}(1-\epsilon)
- [(1-\epsilon_3-\epsilon_4)+ (1-\epsilon_2-\epsilon_4) \omega_3 
 ] ~I^{(4)}_{2,0}(1-\epsilon)
\\
\bar a_{0 0} ~I^{(4)}_{3,0}(1-\epsilon)
+
b_{0 0} ~I^{(4)}_{3,0}(\epsilon)
=
-(1-\epsilon_1) ~I^{(4)}_{3,1}(1-\epsilon)
- [(1-\epsilon_3-\epsilon_4)+ (1-\epsilon_2-\epsilon_4) \omega_3 
 ] ~I^{(4)}_{3,0}(1-\epsilon)
\end{array}
\right.
\end{align}
where the minus sign in the lhs of the first line is due a careful
treatment of phases.
In the limit $\omega_w \rightarrow \infty$
eq. (\ref{Green-global-constraints-sign+reg-gl}) allows to fix $c_{00}$
and again $\bar a_{0 0}$ as
\begin{align}
\left\{
\begin{array}{c}
\bar a_{0 0} ~I^{(4)}_{2,0}(\epsilon)
-
c_{0 0} ~I^{(4)}_{2,0}(1-\epsilon)
=
-\epsilon_1 ~I^{(4)}_{2,1}(\epsilon)
\\
\bar a_{0 0} ~I^{(4)}_{3,0}(\epsilon)
+
c_{0 0} ~I^{(4)}_{3,0}(1-\epsilon)
=
-\epsilon_1 ~I^{(4)}_{3,1}(\epsilon)
\end{array}
\right.
\end{align}
The two previous ways of fixing $\bar a_{0 0}$ must be compatible and
this can be easily verified at least in the $\omega_3 \rightarrow 1^-$ limit.
 
%%%%%%%%%%%%%%%%%%%%%%%%%%%%%%%%%%%%%%%%%%%%%%%%%%%%%%%%%%%%%%%%%%%%%%
\section{The quantum twists correlators}
\label{sect:quantum_corr}
In this section we want to compute the $N$ twists correlators in the
$N-2$ different sectors determined by $M$.
We can generically write the $N$ twists correlators in the $M$ sector
as
\begin{equation}
\langle \prod_{i=1}^N \sigma_{\epsilon_i, f_i}(x_i) \rangle
=
\frac{
A_{(N, M)}(\omega_{j\ne 1,2,N}) e^{-S_{E,cl}(x_i,\epsilon_i, f_i)}
}{
\prod_{1\le i< j \le N} (x_i -x_j)^{\Delta_{i j}}
}
\label{generic-N-twists}
\end{equation}
The powers $\Delta_{i j}$ can be completely fixed as follows.
From the proper behavior for $x_i\rightarrow \infty$ we get the
constraints
$\sum_{l\ne i} \Delta_{i l}= 2 \Delta( \sigma_{\epsilon_i, f_i}) =
\epsilon_i(1-\epsilon_i)$
where we have defined $\Delta_{j i} = \Delta_{i j}$ for $j> i$.
Now redefining $A \rightarrow 
A~ \prod_{3\le i< j \le N-1} (\omega_i -\omega_j)^{\Delta_{i j}}
~\prod_{3\le i \le N-1} (1- \omega_i)^{\Delta_{2 i}}
~\prod_{3\le i \le N-1} \omega_i^{\Delta_{i N}}
$ and remembering that $\omega_2=1$, $\omega_N=0$ and 
$(\omega_i -\omega_j) \propto (x_i -x_j)$
we can set all $\Delta$s to zero but
$\Delta_{1 i}, \Delta_{1 2}, \Delta_{1 N}$ ($3\le i \le N-1$) and $\Delta_{2 N}$ which
can now be fixed by the first set of constraints.
Therefore we can choose a ``gauge'' where the previous correlator can
be written
\begin{align}
\langle \prod_{i=1}^N \sigma_{\epsilon_i, f_i}(x_i) \rangle
=&
\frac{
1
}{
\prod_{3\le i \le N-1} (x_1 -x_i)^{\epsilon_i(1-\epsilon_i)}
}
\nonumber\\
&\cdot
\frac{
1
}{
(x_1 -x_2)^{ \oh[ \epsilon_1(1-\epsilon_1) -\sum_{i=3}^{N-1} \epsilon_i(1-\epsilon_i)
+\epsilon_2(1-\epsilon_2)-\epsilon_N(1-\epsilon_N)]}
}
\nonumber\\
&\cdot
\frac{1}{
(x_1 -x_N)^{ \oh[ \epsilon_1(1-\epsilon_1) - \sum_{i=3}^{N-1} \epsilon_i(1-\epsilon_i)
+\epsilon_N(1-\epsilon_N)-\epsilon_2(1-\epsilon_2)]}
}
\nonumber\\
&\cdot
\frac{1}{
(x_2 -x_N)^{ \oh[
      \epsilon_2(1-\epsilon_2)+\epsilon_N(1-\epsilon_N)-\epsilon_1(1-\epsilon_1)
  + \sum_{i=3}^{N-1} \epsilon_i(1-\epsilon_i)]}
}
\nonumber\\
&\cdot
A_{(N, M)}(\omega_{j\ne 1,2,N})~ e^{-S_{E,cl}(x_i,\epsilon_i, f_i)}
\label{gauge-fixed-N-twists}
\end{align}

We can now proceed in the usual way. We first compute
the expectation value of the energy-momentum tensor as
\begin{align}
\langle\langle T(z) \rangle\rangle
&=
\frac{
\langle 
\dz \cX_q(z) \dz \cbX_q(z) \sigma_{\epsilon_1,f}(x_1) \dots \sigma_{\epsilon_N,f}(x_N)
\rangle
}{
\langle 
\sigma_{\epsilon_1,f}(x_1) \dots \sigma_{\epsilon_N,f}(x_N)
\rangle
}
\nonumber\\
&=
\lim_{w\rightarrow z} g(z,w) -\frac{1}{(z-w)^2}
\end{align}
then  using the OPE
\begin{equation}
T(z) \sigma_{\epsilon_i,f_i}(x_i)
\sim
\frac{\epsilon_i(1-\epsilon_i)}{(z-x_i)^2}
+
\frac{\partial_{x_i} \sigma_{\epsilon_i,f_i}(x_i)}{z-x_i}
+O(1)
\end{equation}
we compute
\begin{align}
\partial_{x_i}\ln \langle \prod_{i=1}^N \sigma_{\epsilon_i,f}(x_i)\rangle
&=
\lim_{z\rightarrow x_i} 
(z-x_i)\left[ 
\langle\langle T(z) \rangle\rangle
-
\frac{\epsilon_i(1-\epsilon_i)}{(z-x_i)^2}
\right]
\end{align}
The function $A_{(N,M)}$ in the quantum case where $f_i=f$
can be determined from eq. (\ref{generic-N-twists}) ($j\ne 1, 2, N$) as
\begin{align}
\partial_{x_j} \ln \langle \prod_{i=1}^N \sigma_{\epsilon_i,f}(x_i) \rangle
&=
- \sum_{l\ne j} \frac{\Delta_{l j}}{x_j-x_l}
+
\frac{\partial \omega_j}{\partial x_j} 
\frac{\partial \ln A_{(N,M)}}{\partial \omega_j}
\label{eq-fix-A}
\end{align}
%2013-11-11
%The full expression for $A_{(N,M)}$, i.e. the case where all $f_i$
%are generically different,  can be obtained by multiplying the quantum
%value for the classical action as in
%eq. (\ref{general_corr_from_quantum_and_classical}). 

\subsection{The $M_{ccw}=N-1$, $M_{cw}=1$ cases}
Using the expansion (\ref{def-S-P}) for $S$
and the constraints (\ref{S-dS-const})
we can easily deduce that
\begin{align}
\langle\langle T(z) \rangle\rangle
&=
\oh
\left(\frac{\partial\omega_z}{\partial z}\right)^2
\left[
\sum_{j=2}^N \frac{\epsilon_j}{(\omega_z-\omega_j)^2}
-
\left( \sum_{j=2}^N \frac{\epsilon_j}{\omega_z-\omega_j}\right)^2
+
\prod_{j=2}^N \frac{1}{\omega_z-\omega_j}
  \frac{\partial^2 S}{\partial \omega_w^2}\Big|_{\omega_w=\omega_z}
\right]
\label{T-Mcw1}
\end{align}
It then follows that ($j\ne 1 , 2,N$)
\begin{align}
\partial_{x_j} \log \langle  \prod_{i=1}^N \sigma_{\epsilon_i,  f_i=f}(x_i) \rangle
&=
- 
\frac{\epsilon_j}{ x_j-x_1}
\nonumber\\
&
+
\epsilon_j
\left[
-
\sum_{l\ne 1,j} \frac{\epsilon_l}{x_j-x_l}
+
\frac{M-\epsilon_1}{x_j-x_1}
\right]
\nonumber\\
&
+
\oh
\prod_{l\ne 1,j} \frac{1}{\omega_j-\omega_l}
  \frac{\partial \omega_j}{\partial x_j}
  \frac{\partial^2 S}{\partial \omega_w^2}\Big|_{\omega_w=\omega_z=\omega_j}
\label{d-log-prod-s-gen-step0}
\end{align}
from which we can obtain $A_{(N,M)}$ using eq. (\ref{eq-fix-A}) to get
\begin{align}
\frac{\partial \ln A_{(N,M)}}{\partial \omega_j}
&=
\oh
\prod_{l=2; l\ne j}^N \frac{1}{\omega_j-\omega_l}
  \frac{\partial^2 S}{\partial \omega_w^2}\Big|_{\omega_w=\omega_z=\omega_j}
+
\sum_{l=2; l\ne j}^N \frac{\Delta_{j l} - \epsilon_j \epsilon_l
}{\omega_j-\omega_l}
\end{align}
The main issue is then to compute 
$  \frac{\partial^2 S}{\partial  \omega_w^2}\Big|_{\omega_w=\omega_z}$. 
This can be done immediately in two cases, i.e.
$M_{cw}=1$  for which $\frac{\partial^2 S}{\partial  \omega_w^2}=0$
since the maximum $\omega_w$ power is 1
and $M_{ccw}=N-1$ for which $\frac{\partial^2 S}{\partial  \omega_z^2}=0$
since the maximum $\omega_z$ power is 1 as it is obvious from eq. (\ref{def-S-P}).
In the former case we get
\begin{align}
\langle \prod_{i=1}^N \sigma_{\epsilon_i, f_i=f}(x_i) \rangle
=&
C_{(N,M=1)}(\epsilon)~
\frac{
\prod_{3\le j<l \le N-1} (\omega_j-\omega_l)^{ 
% general :
%\oh [(1-\epsilon_j)(1-\epsilon_l) + \epsilon_j \epsilon_l]
% this case M=1 :
  -\epsilon_j\epsilon_l
}
}{
\prod_{3\le i \le N-1} (x_1 -x_i)^{\epsilon_i(1-\epsilon_i)}
}
\nonumber\\
&\cdot
\frac{
\prod_{3\le l\le N-1} (1-\omega_l)^{ 
% general :
%\oh [    (1-\epsilon_2)(1-\epsilon_l) + \epsilon_2 \epsilon_l]}
% M=1 :
  -\epsilon_2\epsilon_l
}
}{
(x_1 -x_2)^{ \oh[ \epsilon_1(1-\epsilon_1) -\sum_{i=3}^{N-1} \epsilon_i(1-\epsilon_i)
+\epsilon_2(1-\epsilon_2)-\epsilon_N(1-\epsilon_N)]}
}
\nonumber\\
&\cdot
\frac{
\prod_{3\le j\le N-1} \omega_j^{ 
% general:
%\oh [    (1-\epsilon_j)(1-\epsilon_N) + \epsilon_j \epsilon_N]
% M=1 :
  -\epsilon_j\epsilon_N
}
}{
(x_1 -x_N)^{ \oh[ \epsilon_1(1-\epsilon_1) -\sum_{i=3}^{N-1} \epsilon_i(1-\epsilon_i)
+\epsilon_N(1-\epsilon_N)-\epsilon_2(1-\epsilon_2)]}
}
\nonumber\\
&\cdot
\frac{
1
}{
(x_2 -x_N)^{ \oh[
      \epsilon_2(1-\epsilon_2)+\epsilon_N(1-\epsilon_N)-\epsilon_1(1-\epsilon_1)
+\sum_{i=3}^{N-1} \epsilon_i(1-\epsilon_i)
]}
}
\label{N_M1-quantum-sigmas-correlator}
\end{align}
while in the latter we get the same result but with the substitution
$\epsilon \rightarrow 1 -\epsilon$ by expanding  $\omega_z$ around $\omega_w$.
The coefficients $C_{N,1}$ and $C_{N,N-1}$ will be fixed in section 
\ref{sub:normalizations} and are given in eq.s (\ref{final_C_norm}).

%\subsection{The $M_{ccw}=N-2$ $M_{cw}=2$ case}

\subsection{The $N\ge 4 $ and  $ N-2 \ge M \ge 2$ cases}
For all the other cases it is enough to use a slight modification of
the  technique used in \cite{Atick:1987kd} (see also \cite{Abel:2003yx}).
\COMMENTO{
SBAGLIATO!
Nevertheless a more careful evaluation of the approach we discuss 
in this section while developing the necessary steps 
leads to the constraints quoted
in the title of this section while all the other cases are then
obtained by contemporary reduction of $N$ and $M$ as discussed in
section (\ref{sub:N-1-from-N}).
}

The main idea of this approach is to 
define a new basis for the classical solutions (see eq.s (\ref{class-sols})) 
and consequently for the
non singular part of the derivative of the Green function $g(z,w)$ 
(see eq. (\ref{g-gs-gr}))  
which are closed under certain operations needed to compute the correlators.
  
We start therefore by defining a new basis for the classical solutions 
\begin{align}
\dz_\omega \cX^{(I)}(\omega)
&=
\prod_{j=2}^N (\omega-\omega_j)^{\epsilon_j-1}
\prod_{l\in S_I} (\omega-\omega_l)
~~~~
I\in S
\nonumber\\
\dz_\omega \cbX^{(\bar I)}(\omega)
&=
\prod_{j=2}^N (\omega-\omega_j)^{-\epsilon_j}
\prod_{l\in \bar S_{\bar I}} (\omega-\omega_l)
~~~~
\bar I\in \bar S
\end{align}
where we have defined two ordered sets 
\begin{equation}
S=\{ N-M-1 ~\mbox{arbitrary different  indexes chosen among}~  3,\dots
N-1 \}
\end{equation}
and 
\begin{equation}
\bar S=\{ M-1 ~\mbox{arbitrary different  indexes chosen among}~  3,\dots
N-1 \}
\end{equation}
and the subsets $S_I=S-\{ I\}$  for any $I\in S$ and similarly for $\bar
S_{\bar I}$.
In order to be able to define the previous basis as a linear combination
of the original one (\ref{basis-class-sols}) we need  that
both $n \ge 0$  and $ r \ge 0$, i.e $N-2 \ge M \ge 2$.
In particular what follows works even if either $S_J=\emptyset$ 
or $\bS_\bJ=\emptyset$, i.e. $M=N-2$ or $M=2$ for example when $N=4$ and $M=2$.
 
We can now expand the regular part of $g$ and $h$, $l$ as
\begin{align}
g_{r (N,M)}(z, w; \{x_i\} )
&=
\frac{\partial \omega_z}{\partial z}
\frac{\partial \omega_w}{\partial w}
\sum_{I\in S}\sum_{\bI \in \bS} \bar a_{I \bI}(\omega_j) 
~\dz_\omega \cX^{(I)}(\omega_z)
~\dz_\omega \cbX^{(\bI)}(\omega_w)
\nonumber\\
h_{ (N,M)}(z, w; \{x_i\} )
&=
\frac{\partial \omega_z}{\partial z}
\frac{\partial \omega_w}{\partial w}
\sum_{\bI\in \bS}\sum_{\bJ \in \bS} b_{\bI \bJ}(\omega_j) 
~\dz_\omega \cbX^{(\bI)}(\omega_z)
~\dz_\omega \cbX^{(\bJ)}(\omega_w)
\nonumber\\
l_{(N,M)}(z, w; \{x_i\} )
&=
\frac{\partial \omega_z}{\partial z}
\frac{\partial \omega_w}{\partial w}
\sum_{I\in S}\sum_{J \in S} c_{I J}(\omega_j) 
~\dz_\omega \cX^{(I)}(\omega_z)
~\dz_\omega \cX^{(J)}(\omega_w)
\end{align}
Then we can find a solution of the first of the constraints 
in eq. (\ref{glob-consts-gh-gl}) as
\begin{align}
g_{(N,M)}(z, w; \{x_i\} )
&=
g_{s (N,M)}(z, w; \{x_i\} )
-
\frac{\partial \omega_z}{\partial z}
\sum_{i=1}^{N-2}
\sum_{I\in S}
(W^{-1})^i_I 
\dz_\omega \cX^{(I)}(\omega_z)
\int_{\cI_i} \frac{d \omega}{ (\omega-\omega_w)^2} P S_0(\omega,\omega_w)
\nonumber\\
&=
g_{s (N,M)}(z, w; \{x_i\} )
-
\frac{\partial \omega_w}{\partial w}
\sum_{i=1}^{N-2}
\sum_{\bI\in \bS}
(W^{-1})^i_\bI 
\dz_\omega \cbX^{(\bI)}(\omega_w)
\int_{\cI_i} \frac{d \omega}{ (\omega_z-\omega)^2} P S_0(\omega_z,\omega)
\end{align}
where $S_0(\omega_z,\omega_w)$ is the same as in the second equation
in (\ref{def-S-P})  but for the singular part of $g$, i.e. with
coefficients $a_{(0)}$ as in eq. (\ref{g-gs-gr}).
Moreover we have also defined the  $(N-2)\times (N-2)$ matrix $W$ as\footnote{
Again as in eq.s (\ref{glob-consts-gh-gl}) it is important that the
integration is once above and once below the cut as this  ensures that
both integrals have the same phase modulus $\pi$.
}
\begin{align}
W_{i}^{~I}
&=
\int_{\omega_{i+2}}^{\omega_{i+1}} d \omega ~\partial_\omega\cX^{(I)}(\omega+i 0^+)
~~~~i=1,\dots N-2,~~~I\in S
\nonumber\\
W_{i}^{~\bI}
&=
\int_{\omega_{i+2}}^{\omega_{i+1}} d \omega ~\partial_\omega\cbX^{(I)}(\omega-i 0^+)
~~~~i=1,\dots N-2,~~~\bI\in \bS
\label{W-integrals}
\end{align}
From this expression is immediate to compute the energy-momentum
tensor expectation value which can be split into a singular part as in
eq. (\ref{T-Mcw1}) but with the substitution $S\rightarrow S_0$ and 
a regular part as
\begin{align}
\langle\langle T_{r}(w) \rangle\rangle
&=
-
\frac{\partial \omega_w}{\partial w}
\sum_{i=1}^{N-2}
\sum_{I\in S}
(W^{-1})^i_I 
\int_{\cI_i} \frac{d \omega}{ (\omega-\omega_w)^2} 
\frac{
\prod_{j=2}^N (\omega-\omega_j)^{\epsilon_j-1}
}{
\prod_{j\not\in S_I} (\omega_w-\omega_j)
}
S_0(\omega,\omega_w)
\end{align}
where $j\not\in S_I$ means $j\in \{2, \dots N\} \setminus S_I$.
If we consider $J\in S %%%%%%%%%\cap \bS
$ 
we can then evaluate\footnote{
There are two integrals which are actually divergent but their sum is
however convergent. These integrals correspond to the intervals for
which $\omega_J$ is a boundary point.
}
\footnote{
We restrict to the case $J\in S$ because otherwise
eq. (\ref{d-log-prod-s-step1}) would contain the sum over all $I\in S$ and 
eq. (\ref{integrand-expansion}) would contain the sum over all
possible $\partial_{\omega_j} \dz_\omega \cX^{(I)}(\omega)$ each with
a non trivial coefficient.
}
\begin{align}
\lim_{w \rightarrow x_J}
&(w-x_J)
\langle\langle T_{r}(w) \rangle\rangle
\nonumber\\
&=
-
\frac{\partial \omega_J}{\partial x_J}
\sum_{i=1}^{N-2}
(W^{-1})^i_J
\int_{\cI_i} \frac{d \omega}{ (\omega-\omega_J)^{3-\epsilon_J}} 
\frac{
\prod_{j=2; j\ne J}^N (\omega-\omega_j)^{\epsilon_j-1}
}{
\prod_{j\not\in S} (\omega_J-\omega_j)
}
S_0(\omega,\omega_J)
\label{d-log-prod-s-step1}
\end{align}
\COMMENTOOK{Warning regularization}
Now following \cite{Atick:1987kd} we rewrite the integrand as
\begin{align}
\frac{1}{ (\omega-\omega_J)^{3-\epsilon_J}} 
\frac{
\prod_{j=2; j\ne J}^N (\omega-\omega_j)^{\epsilon_j-1}
}{
\prod_{j\not\in S} (\omega_J-\omega_j)
}
S_0(\omega,\omega_J)
&=
\partial_{\omega_J} \dz_\omega \cX^{(J)}(\omega)
+ \sum_{L\in S} T^J_L \dz_\omega \cX^{(L)}(\omega)
\label{integrand-expansion}
\end{align}
\COMMENTOOK{This is possible only for $J\in S$ since otherwise we get
a sum of  all possible $\partial_{\omega_j} \dz_\omega \cX^{(I)}(\omega)$. }.
The leading singularity is
$O\left( (\omega-\omega_J)^{-2+\epsilon_J} \right)$ because
$S_0(\omega_J,\omega_J)=0$  as follows from the first equation in
(\ref{S-dS-const}) 
when evaluated for $\omega_w=\omega_J$.
Moreover when the left hand side is subtracted the leading singularity
and multiplied for 
$ \prod_{j=2}^N (\omega-\omega_j)^{1-\epsilon_j}/ \prod_{j\in S} (\omega-\omega_j) $
we are left with a rational function with poles at $\omega_I$ ($I\in
S$) which vanish at $\omega=\infty$ as the right hand side.
Because of the sum over $i$ in eq. (\ref{d-log-prod-s-step1}) the only $T^J_L$ needed is
\begin{align}
T^J_J
&=
-(1-\epsilon_J) \sum_{l\in S_J} \frac{1}{\omega_J-\omega_l}
+
\oh
\prod_{l\in S_J} \frac{1}{\omega_J-\omega_l}
\prod_{l\not\in S} \frac{1}{\omega_J-\omega_l}
\partial^2_{\omega_z} S_0(\omega_J, \omega_J)
\end{align}
When we insert this value into eq. (\ref{d-log-prod-s-step1}) and add
the contribution from the singular part which has the same expression as
eq. (\ref{d-log-prod-s-gen-step0}) with $S\rightarrow S_0$ 
and $\epsilon\rightarrow 1-\epsilon$ since we have here $\partial^2_{\omega_z} S_0$
we get
\begin{align}
\partial_{x_J} \log \langle  \prod_{i=1}^N \sigma_{\epsilon_i,  f_i=f}(x_i) \rangle
&=
- 
\frac{(1-\epsilon_J) (2-\epsilon_J)}{ x_J-x_1}
\nonumber\\
&
-
2(1-\epsilon_J)
\left[
\sum_{l\ne 1,J} \frac{1-\epsilon_l}{x_J-x_l}
+
\frac{M-N+1-\epsilon_1}{x_J-x_1}
\right]
\nonumber\\
&-
\frac{\partial \omega_J}{\partial x_J}
\left[
\sum_{i=1}^{N-2} (W^{-1})^i_J  \partial_{\omega_J}W^J_i
-(1-\epsilon_J) \sum_{l\in S_J} \frac{1}{\omega_J-\omega_l}
\right]
\label{d-log-prod-sigma-d2z}
\end{align}
from which the dependence on $S_0$ has disappeared but we are left with
a dependence on $\partial_{\omega_J}W^J_i$. 
Differently from what done in  \cite{Atick:1987kd} 
%(see also \cite{Abel:2003yx}) 
we cannot rely on fact that twists have both an
holomorphic and antiholomorphic dependence in order to end the
computation using 
\begin{align}
\det W
&=
\sum_{i=1}^{N-2}\left[
\sum_{I\in S}  (W^{-1})^i_I  \partial_{\omega_J}W^I_i
+
\sum_{\bI\in \bS}  (W^{-1})^i_\bI  \partial_{\omega_J}W^\bI_i
\right]
\end{align}
and 
\begin{equation}
\partial_{\omega_J} W^{ I\ne J}_i
=
\frac{\epsilon_I}{\omega_J-\omega_I}
\left(
 W^{ J}_i
-
 W^{ I}_i
\right)
\label{dWI_ne_J}
\end{equation}
%Moreover otherwise $\partial_{\omega_J} W^{\bI\ne J}_i$ cannot
%expanded on $\partial_\omega \cbX^{(\bI)}$
%because $J\in \bS$ and therefore $\partial_{\omega_J} W^{\bJ=J}_i$         
%does not vanish.
Instead we have to rely on the second of the
(\ref{glob-consts-gh-gl}) constraints (or better its complex conjugate
which is has the same expression with the substitution $\pm i0^+
\rightarrow \mp i 0^+$).
Analogously as before we require $\bJ \in \bS$ and we get
\begin{align}
\partial_{x_\bJ} \log \langle  \prod_{i=1}^N \sigma_{\epsilon_i,  f_i=f}(x_i) \rangle
&=
- 
\frac{\epsilon_\bJ (1+\epsilon_\bJ)}{ x_\bJ-x_1}
\nonumber\\
&
-
2 \epsilon_\bJ
\left[
\sum_{l\ne 1,\bJ} \frac{\epsilon_l}{x_\bJ-x_l}
+
\frac{-M+\epsilon_1}{x_\bJ-x_1}
\right]
\nonumber\\
&-
\frac{\partial \omega_\bJ}{\partial x_\bJ}
\left[
\sum_{i=1}^{N-2} (W^{-1})^i_\bJ  \partial_{\omega_J}W^\bJ_i
-\epsilon_\bJ \sum_{l\in S_\bJ} \frac{1}{\omega_\bJ-\omega_l}
\right]
\label{d-log-prod-sigma-d2w}
\end{align}
then only if $J=\bJ \in S \cap \bS$ 
we can average the previous expressions 
(\ref{d-log-prod-sigma-d2z}) and (\ref{d-log-prod-sigma-d2w})
Into this average we can use the analogous expression of eq. (\ref{dWI_ne_J})
\begin{equation}
\partial_{\omega_J} W^{ \bI\ne \bJ}_i
=
\frac{1-\epsilon_\bI}{\omega_\bJ-\omega_\bI}
\left(
 W^{\bJ}_i
-
 W^{ \bI}_i
\right)
\end{equation}
and
\begin{align}
\sum_{i=1}^{N-2} (W^{-1})^i_J  \partial_{\omega_J}W^J_i
&+
\sum_{i=1}^{N-2} (W^{-1})^i_\bJ  \partial_{\omega_J}W^\bJ_i
\nonumber\\
&=
\partial_{\omega_J} \det W
-
\sum_{I\in S_J}
\sum_{i=1}^{N-2} (W^{-1})^i_I  \partial_{\omega_J}W^I_i
-
\sum_{\bI \in \bS_J}
\sum_{i=1}^{N-2} (W^{-1})^i_\bI  \partial_{\omega_J}W^\bI_i
\nonumber\\
&=
\partial_{\omega_J} \det W
+
\sum_{I\in S_J} \frac{\epsilon_l}{\omega_J-\omega_l}
+
\sum_{\bI\in \bS_J} \frac{1-\epsilon_l}{\omega_J-\omega_l}
\end{align}
to get finally
\begin{align}
\partial_{\omega_J} \log A_{(N,M)}(\omega_j)
=
\partial_{\omega_J} \log
\Big[
(\det W)^{-\oh}
&
\prod_{l\in S_J} (\omega_J -\omega_l)^\oh
\prod_{l\in \bS_J} (\omega_J -\omega_l)^\oh
\nonumber\\
&
\prod_{l\ne 1,J} (\omega_J-\omega_l)^{\Delta_{J l} 
 - \oh [    (1-\epsilon_J)(1-\epsilon_l)
+ \epsilon_J \epsilon_l]}
\Big]
\end{align}
The previous equation is valid only for $J\in S \cap \bS$ 
but if, 
\COMMENTOOK{
Under the hypothesis that .... the previous equation can be checked by
changing the remaining element.
}
in either $S$ or in $\bS$ there is at least one further element
than those contained in $S \cap \bS$
or if  $S \cap \bS$ contains all the independent $\omega_j$ as in the
$N=4$ case, we can deduce that\footnote{
In the expression we have used $ord(\bI)$ to indicate the order of
  $I$ in the ordered set $S$.
}
\begin{align}
A_{(N,M)}(\omega_j)
&=
\mbox{const}
(\det W)^{-\oh}
\prod_{ord(I)<ord(J);I,J\in S} (\omega_I -\omega_J)^\oh
\prod_{ord(\bI)<ord(\bJ);\bI,\bJ\in \bS} (\omega_\bI -\omega_\bJ)^\oh
\nonumber\\
&~~
\prod_{2\le j<l \le N} (\omega_j-\omega_l)^{\Delta_{j l} 
 - \oh [   (1-\epsilon_j)(1-\epsilon_l) + \epsilon_j \epsilon_l]}
\end{align}
as a consequence of the independence of the result under a change of
the elements of $S$ and/or $\bS$.
%%%%%%%%%%%%%%%%%% COMMENTO
\COMMENTO{
Therefore the previous expression can be deduced only when either 
$N-M-1 > M-1$ or $M-1 > N-M-1$ and this explains the further
constraints on $N$ and $M$ in this section title.
ovvio o a>b o b>a...
}
%%%%%%%%%%%%%%%%%%
Under the change $S\rightarrow S'= (S\setminus \{I_0\}) \cup \{I_1\}$ 
the integrals $ W^I_{(S) i}$ in eq.s (\ref{W-integrals}) transform as
\begin{align}
W^L_{(S') i}
&=
\frac{\omega_{I_1}-\omega_L}{\omega_{I_0}-\omega_L} W^L_{(S) i}
+
\frac{\omega_{I_1}-\omega_{I_0}}{\omega_L-\omega_{I_0}} W^{I_0}_{(S) i}
~~~~ L\ne I_0
\nonumber\\
W^{I_1}_{(S') i}
&=
W^{I_0}_{(S) i}
\end{align}
so that the transformation of the determinant
\begin{align}
\det W_{S',\bS}
&=
\det W_{S,\bS}
\prod_{L\in S_{I_0}} \frac{\omega_{I_1}-\omega_L}{\omega_{I_0}-\omega_L}
\end{align}
is what is needed to compensate the change
$\prod_{ord(I)<ord(J);I,J\in S'} (\omega_I -\omega_J)^\oh
\rightarrow
\prod_{ord(I)<ord(J);I,J\in S} (\omega_I -\omega_J)^\oh$.

The final expression for the $N$ twists correlator in the $M$ sector
is then
\begin{align}
\langle \prod_{i=1}^N \sigma_{\epsilon_i, f_i=f}(x_i) \rangle
=&
C_{(N,M)}(\epsilon)~
\frac{
\prod_{3\le j<l \le N-1} (\omega_j-\omega_l)^{ 
  -\oh [(1-\epsilon_j)(1-\epsilon_l) + \epsilon_j \epsilon_l]}
}{
\prod_{3\le i \le N-1} (x_1 -x_i)^{\epsilon_i(1-\epsilon_i)}
}
\nonumber\\
&\cdot
\frac{
\prod_{3\le l\le N-1} (1-\omega_l)^{ 
  -\oh [(1-\epsilon_2)(1-\epsilon_l) + \epsilon_2 \epsilon_l]}
}{
(x_1 -x_2)^{ \oh[ \epsilon_1(1-\epsilon_1) -\sum_{i=3}^{N-1} \epsilon_i(1-\epsilon_i)
  +\epsilon_2(1-\epsilon_2)-\epsilon_N(1-\epsilon_N)]}
}
\nonumber\\
&\cdot
\frac{
\prod_{3\le j\le N-1} \omega_j^{ 
  -\oh [ (1-\epsilon_j)(1-\epsilon_N) + \epsilon_j \epsilon_N]}
}{
(x_1 -x_N)^{ \oh[ \epsilon_1(1-\epsilon_1) -\sum_{i=3}^{N-1} \epsilon_i(1-\epsilon_i)
  +\epsilon_N(1-\epsilon_N)-\epsilon_2(1-\epsilon_2)]}
}
\nonumber\\
&\cdot
\frac{
1
}{
(x_2 -x_N)^{ 
  \oh[ \epsilon_2(1-\epsilon_2)+\epsilon_N(1-\epsilon_N)-\epsilon_1(1-\epsilon_1)
  +\sum_{i=3}^{N-1} \epsilon_i(1-\epsilon_i)
]}
}
\nonumber\\
&\cdot
(\det W_{S,\bS})^{-\oh}
\prod_{ord(I)<ord(J);I,J\in S} (\omega_I -\omega_J)^\oh
\prod_{ord(\bI)<ord(\bJ);\bI,\bJ\in \bS} (\omega_\bI -\omega_\bJ)^\oh
\label{N-quantum-sigmas-correlator}
\end{align}
Notice that the previous expression is true even if there is only
one element in $\bS$ in which case the product
$\prod %_{ord(\bI)<ord(\bJ);\bI,\bJ\in \bS}  
(\omega_\bI -\omega_\bJ)^\oh$ is simply $1$.
Similarly for the $S$ case.
In subsection \ref{sub:normalizations} 
we will fix the constant $C_{(N,M)}(\epsilon)$.

\subsection{$N-1$ amplitudes from $N$ amplitudes}
\label{sub:N-1-from-N}
%The results of the previous sections apply strictly speaking only for
%$N\ge 4$ and now we would like to show that they also apply for the
%other cases.
We want to check the consistency of the results of the previous section.
We do this by making $x_{j+1}$ coalesce with $x_j$ and so 
deducing the $N-1$ twists correlators from $N$ twists
ones.

We start noticing that from the $(N,M)$ sector we can generically compute both
$(\tilde N, \tilde M)=(N-1,M)$ and $(\tilde N, \tilde M)=(N-1, M-1)$ 
sectors depending whether 
$\epsilon_j+\epsilon_{j+1}<1$ or $\epsilon_j+\epsilon_{j+1}>1$.
Exceptions are the $M=1$ case where only $\tilde M=1$ is
possible and  $M=N-1$ where  only $\tilde M=\tilde N-1=N-2$ is
possible.

\subsubsection{$(N,1)$ into $(N-1,1)$ case}
Starting from eq. (\ref{N_M1-quantum-sigmas-correlator}) we can very
easily take the limit $x_{J+1} \rightarrow x_J$.
When we use 
\begin{equation}
\epsilon_J(1-\epsilon_J)+
\epsilon_{J+1}(1-\epsilon_{J+1})
=
\tilde \epsilon_J(1- \tilde \epsilon_J)
+ 2\epsilon_J \epsilon_{J+1}
~~~~
\tilde \epsilon_J = \epsilon_J +\epsilon_{J+1}
\end{equation}
and
\begin{equation}
\omega_J - \omega_{J+1}
= 
\frac{
(x_{J+1} - x_J) ( x_N -x_1) 
}{
(x_{J} - x_1) ( x_{J+1} -x_1) 
}
\frac{x_2 - x_1 }{x_2 - x_N}
\end{equation}
we find 
\begin{align}
\langle \prod_{i=1}^N \sigma_{\epsilon_i, f}(x_i) \rangle
\sim_{x_{J+1} \rightarrow x_J}
(x_J - x_{J+1})^{ -\epsilon_J \epsilon_{J+1}}
~\cM(\epsilon_J, \epsilon_{J+1})
~\langle \prod_{\tilde i=1}^{N-1} 
\sigma_{\tilde \epsilon_{\tilde i},  f}(x_{\tilde i}) \rangle
\label{N_1-into-N-1_1}
\end{align}
and the consistency relation for the normalizations
\begin{equation}
C_{(N,1)}(\epsilon) 
=
C_{(N-1,1)}(\tilde \epsilon)
~
\cM(\epsilon_J, \epsilon_{J+1})
\label{C-C-M1}
\end{equation}
where $\tilde \epsilon$ are the twists of the $(N-1,1)$ theory defined by
$\tilde \epsilon_j =\epsilon_j$ for $j<J$,
$\tilde \epsilon_J =\epsilon_J+ \epsilon_{J+1}$ for $j=J$ and
$\tilde \epsilon_j =\epsilon_{j+1}$ for $j>J$.
Actually all the previous equations work even when we consider the
$\omega_j\rightarrow \infty$ ($x_j\rightarrow x_1$) limit.

\subsubsection{$(N,N-1)$ into $(N-1,N-2)$ case}
In a way completely analogous to that done in the previous subsection
we get
\begin{align}
\langle \prod_{i=1}^N \sigma_{\epsilon_i, f}(x_i) \rangle
\sim_{x_{J+1} \rightarrow x_J}
(x_J - x_{J+1})^{ -(1-\epsilon_J) (1-\epsilon_{J+1})}
~\cN(\epsilon_J, \epsilon_{J+1})
~\langle \prod_{\tilde i=1}^{N-1} 
\sigma_{\tilde \epsilon_{\tilde i},  f}(x_{\tilde i}) \rangle
\label{N_N-1-into-N-1_N-2}
\end{align}
and the consistency relation the consistency relation for the
normalization coefficients
\begin{equation}
C_{(N,1)}(\epsilon) 
=
C_{(N-1,1)}(\tilde \epsilon)
~
\cN(\epsilon_J, \epsilon_{J+1})
\label{C-C-MN-1}
\end{equation}
where $\tilde \epsilon$ are the twists of the $(N-1,1)$ theory defined by
$\tilde \epsilon_j =\epsilon_j$ for $j<J$,
$\tilde \epsilon_J =\epsilon_J+ \epsilon_{J+1}-1$ for $j=J$ and
$\tilde \epsilon_j =\epsilon_{j+1}$ for $j>J$.
Again all works in the 
$\omega_j\rightarrow \infty$ ($x_j\rightarrow x_1$) limit.

\subsubsection{$(N,M)$ into $(N-1,M)$ with $2\le M\le N-2$ case}
In this case we start from the general expression
(\ref{N-quantum-sigmas-correlator}) and 
choose the sets $S$ and $\bS$ so that $J\in S$,
$J\not \in \bS$ and $J+1 \not\in S$ then we now show that 
the new sets $\tilde S$
and $\tilde \bS$ are given by $\tilde S= S_J= S\setminus \{J\}$  and
$\tilde \bS = \bS$.
%Because of this setup what follows applies to $M\ge 2$ only.

The previous choices are dictated by the need of having a simple and
clean way of computing the limit of $\det W_{S,\bS}$.
In particular while the interval $[\omega_{J+1}, \omega_J]$ vanishes
$\partial \cX^{(I\ne J)}_S$ and $\partial \cbX^{(\bI)}_\bS$ become the new
$\partial \cX^{(I\ne J)}_{\tilde S}$ and $\partial \cbX^{(\bI)}_{\tilde \bS}$
and 
$\partial \cX^{(J)}$ develops a not integrable singularity at
$\omega_J= \omega_{J+1}$ and gives the leading singularity of $\det
W_{S,\bS}$, explicitly we find\footnote{See appendix
  \ref{app:details_reductions}
for an example of the computations involved in the special case $N=4$ $M=2$.
}
\begin{align}
\det W_{S,\bS}
\sim
W^{(J)}_{(S,\bS) i=J-1}
\det W_{\tilde S, \tilde \bS}
\end{align}
with
\begin{align}
W^{(J)}_{(S,\bS) i=J-1}
\sim
(\omega_J -\omega_{J+1})^{1 -\epsilon_J -\epsilon_{J+1}}
~
e^{-i \pi \epsilon_J} ~B(\epsilon_J, \epsilon_{J+1})
~
\prod_{l\ne 1, J, J+1} (\omega_J -\omega_l)^{-\epsilon_l}
 \prod_{L\in S_J} (\omega_J -\omega_L)
\end{align}
where $B(\cdot, \cdot)$ is Euler Beta function.
Using these results into (\ref{N-quantum-sigmas-correlator}) with a
not so short computation we find the expected result
\begin{align}
\langle \prod_{i=1}^N \sigma_{\epsilon_i, f_i=f}(x_i) \rangle
\sim_{x_{J+1} \rightarrow x_J}
(x_J - x_{J+1})^{ -\epsilon_J \epsilon_{J+1}}
~\cM(\epsilon_J, \epsilon_{J+1})
~\langle \prod_{i=1, i\ne J}^N \sigma_{\epsilon_i, f_i=f}(x_i) \rangle
\label{N_M-into-N-1_M}
\end{align}
and a relation among the amplitude normalizations 
and the OPE normalization in eq. (\ref{sigma-sigma-OPE0}) 
which up to a phase reads
\begin{equation}
C_{(N,M)}(\epsilon) 
~[ B(\epsilon_J, \epsilon_{J+1}) ]^{-\oh}
=
C_{(N-1,M)}(\tilde \epsilon)
~
\cM(\epsilon_J, \epsilon_{J+1})
\label{C-C-cM}
\end{equation}
where $\tilde \epsilon$ are the twists of the $(N-1,M)$ theory, i.e.
$\tilde \epsilon_j =\epsilon_j$ for $j<J$,
$\tilde \epsilon_J =\epsilon_J+ \epsilon_{J+1}$ for $j=J$ and
$\tilde \epsilon_j =\epsilon_{j+1}$ for $j>J$.

It is worth noticing that the previous result (\ref{N_M-into-N-1_M}) shows that 
eq. (\ref{N-quantum-sigmas-correlator}) is valid even when $S$ has
only one element. 
%and $\bS$ at least two as stated just below
%eq. (\ref{N-quantum-sigmas-correlator}).
If we perform the reduction from this case, i.e. with $(N+1,N-1)$
and we compare with the expression for the $(N, N-1)$ amplitudes  
%obtained reducing and the one computed directly similarly
%to the one in eq. (\ref{N_M1-quantum-sigmas-correlator}) 
we deduce that
\begin{equation}
\det W_{(\emptyset, \bar S)} 
\prod_{ord(\bI)<ord(\bJ); \bI,\bJ\in \bar S} (\omega_\bI -\omega_\bJ)^{-1}
\propto 
\prod_{2 \le j < l \le N} (\omega_j -\omega_l)^{\epsilon_j+ \epsilon_l-1}
\end{equation}
where $W_{(\emptyset, \bar S)} $ is simply the matrix $\parallel
W_i^\bI \parallel$. In other words certain determinants of order $N-2$
($card(\bar S)=M-1=N-2$) of  Lauricella hypergeometric functions of
order $N-3$ (since all $ W_i^\bI $ can be expressed using $I^{(N)}$)
are a product of powers. 
This could point to that also the general $\det
W_{S,\bS}$ may be expressed as an elementary function.

For the special case where both $S$ and $\bS$ have just one element,
i.e. for $N=4$, $M=2$ a direct and little different computation is
needed but the result is the same.

For checking the consistency of the approach and of the normalization
coefficients we determine in the next section it is worth considering
the $\omega_J \rightarrow \infty$ limit. The result for the
normalization coefficients in this case
 is based on the relation
\begin{align}
\det W_{S,\bS}
\sim
&
W^{(J)}_{(S,\bS) i=J-1}
~\omega_J^{\epsilon_J(N-2 M -1)}
~\det W_{\tilde S, \tilde \bS}
\nonumber\\
\sim
&
B(\epsilon_J,1-\epsilon_1-\epsilon_J) 
~\omega_J^{\epsilon_J(N-2 M -1)-\epsilon_1}
~\det W_{\tilde S, \tilde \bS}
\end{align}
and reads
\begin{equation}
C_{(N,M)}(\epsilon) 
~[ B(\epsilon_J, 1-\epsilon_1-\epsilon_{J}) ]^{-\oh}
=
C_{(N-1,M)}(\tilde \epsilon)
~
\cM(\epsilon_J, \epsilon_{1})
\label{C-C-cM-inf}
\end{equation}
with the new twists given by $\tilde \epsilon_1 =\epsilon_1+ \epsilon_{J}$,
$\tilde \epsilon_j =\epsilon_j$ for $1<j<J$
and
$\tilde \epsilon_j =\epsilon_{j+1}$ for $j>J$.

\subsubsection{$(N,M)$ into $(N-1,M-1)$ case}
In this case we can choose the sets $S$ and $\bS$ so that $\bJ\in \bS$,
$\bJ\not \in S$ and $\bJ+1 \not\in \bS$ then it is possible to
show as in the previous case that 
the new sets $\tilde S$ and $\tilde \bS$ are given by $\tilde S= S$  and
$\tilde \bS = \bS_J= \bS\setminus \{J\}$.
%Again because of this setup what follows applies to $M\ge 3$ only.

In particular it is possible to find analogously as before that the
determinant behaves in the $x_\bJ \rightarrow x_{\bJ+1}$ limit as
\begin{align}
\det W_{S,\bS}
\sim
W^{(\bJ)}_{(S,\bS) i=\bJ-1}
\det W_{\tilde S, \tilde \bS}
\end{align}
and the amplitude reduction gives
\begin{align}
\langle \prod_{i=1}^N \sigma_{\epsilon_i, f_i=f}(x_i) \rangle
\sim
(x_J - x_{J+1})^{- (1-\epsilon_J)(1-\epsilon_{J+1})}
\langle \prod_{i=1, i\ne J}^N \sigma_{\epsilon_i, f_i=f}(x_i) \rangle
\label{N_M-into-N-1_M-1}
\end{align}
It follows a relation among the amplitude normalizations 
and the OPE normalization in eq. (\ref{sigma-sigma-OPE0}) 
which up to a phase reads
\begin{equation}
C_{(N,M)}(\epsilon) 
~[ B(1-\epsilon_J, 1-\epsilon_{J+1}) ]^{-\oh}
=
C_{(N-1,M-1)}(\tilde \epsilon)
~
\cN(\epsilon_J, \epsilon_{J+1})
\label{C-C-cN}
\end{equation}
where $\tilde \epsilon$ are the twists of the $(N-1,M-1)$ theory, i.e.
$\tilde \epsilon_j =\epsilon_j$ for $j<J$,
$\tilde \epsilon_J =\epsilon_J+ \epsilon_{J+1}-1$ for $j=J$ and
$\tilde \epsilon_j =\epsilon_{j+1}$ for $j>J$.

\COMMENTOO{
Again the previous result (\ref{N_M-into-N-1_M-1}) shows that 
eq. (\ref{N-quantum-sigmas-correlator}) is valid even when $\bS$ has
only one element while $S$ has at least two.
}

As in the previous subsection starting from the $(N+1,2)$ amplitude
and reducing it to $(N,1)$ we deduce that
\begin{equation}
\det W_{(S, \emptyset)} 
\prod_{ord(I)<ord(J); I,J\in S} (\omega_I -\omega_J)^{-1}
\propto 
\prod_{2 \le j < l \le N} (\omega_j -\omega_l)^{1-\epsilon_j - \epsilon_l}
\end{equation}
where $W_{(S, \emptyset)} $ is simply the matrix $\parallel W_i^I \parallel$.

%%%%%%%%%%%%%%%% COMMENTO
\COMMENTO{
\subsubsection{ $N=4$ $M=2$ case.}
Starting from the $N=7$ $M=3$ case with 
$S=\{4,5,6\}$ and $\bS=\{3,6\}$
we can first coalesce $3$ and $4$ and reach $N=6$ $M=2$ with 
$S=\{3,5,6\}$ and $\bS=\{6\}$
and then $N=5$ $M=2$ by coalescing $3$ and $5$.
We are then left with 
$S=\{3,6\}$ and $\bS=\{6\}$.
From this point we cannot anymore apply the same steps as before but
we need a direct computation.
}
%%%%%%%%%%%%%%%% COMMENTO

\subsection{Amplitudes and OPEs normalization}
\label{sub:normalizations}
We normalize the 2-point amplitude as
\begin{equation}
\langle \sigma_{\epsilon}(x) \sigma_{1-\epsilon}(y) \rangle
=
\frac{1}{(x-y)^{\epsilon(1-\epsilon)}}
\label{2-twist-norm}
\end{equation}
This normalization is not unique  since any redefinition as 
$\sigma_\epsilon \rightarrow \cR(\epsilon) \sigma_\epsilon $ with
$\cR(\epsilon) ~ \cR(1-\epsilon)=1$ would work.
In particular this kind of redefinition can only be seen in amplitudes
with at least three twist fields since it leaves unchanged amplitudes
involving two twist fields and an arbitrary number of untwisted fields
therefore it cannot be fixed factorizing a 4 twists into an untwisted channel.
If we require the normalizations to be invariant under the symmetry
$\epsilon \leftrightarrow 1-\epsilon$
then all the normalizations are completely fixed (up one constant $k$
and phases) to be
\begin{align}
C_{(N,1)}&= k^{N-2}
\left[ \prod_{j=1}^N
  \frac{ \Gamma(1-\epsilon_j) }{ \Gamma(\epsilon_j) }
\right]^{1/4}
\nonumber\\
C_{(N,M)}&= k^{N-2}
\left[ 
  \frac{\prod_{j=2}^N \Gamma(\epsilon_j) \Gamma(1-\epsilon_j)
}{ \Gamma(\epsilon_1) \Gamma(1-\epsilon_1) }
\right]^{1/4}
~~~~2\le M \le N-2
\nonumber\\
C_{(N,N-1)}&= k^{N-2}
\left[ \prod_{j=1}^N
  \frac{ \Gamma(\epsilon_j) }{ \Gamma(1-\epsilon_j) }
\right]^{1/4}
\label{final_C_norm}
\end{align}
along with the OPE normalizations
\begin{align}
\cM(\alpha,\beta)
&= k
\left[
\frac{\Gamma(1-\alpha)}{\Gamma(\alpha) } 
\frac{\Gamma(1-\beta)}{\Gamma(\beta) } 
\frac{\Gamma(\alpha+\beta)}{\Gamma(1-\alpha-\beta) } 
\right]^{1/4}
\nonumber\\
&= k
\left[
\frac{\Gamma(1-\alpha)}{\Gamma(\alpha) } 
\frac{\Gamma(1-\beta)}{\Gamma(\beta) } 
\frac{\Gamma(1-\gamma)}{\Gamma(\gamma) } 
\right]^{1/4}
~~~~
\alpha+\beta+\gamma=1
\nonumber\\
\cN(\alpha,\beta)
&= k
\left[
\frac{\Gamma(\alpha)}{\Gamma(1-\alpha) } 
\frac{\Gamma(\beta)}{\Gamma(1-\beta) } 
\frac{\Gamma(2-\alpha-\beta)}{\Gamma(\alpha+\beta-1) } 
\right]^{1/4}
\nonumber\\
&= k
\left[
\frac{\Gamma(\alpha)}{\Gamma(1-\alpha) } 
\frac{\Gamma(\beta)}{\Gamma(1-\beta) } 
\frac{\Gamma(\delta)}{\Gamma(1-\delta) } 
\right]^{1/4}
~~~~
\alpha+\beta+\delta=2
\end{align}
which also respect the symmetry $\epsilon \leftrightarrow 1-\epsilon$
as $\cN(\alpha,\beta)=\cM(1-\alpha, 1-\beta)$.
%%%%%%%%%%%%%%%%% COMMENTO %%%%%%%%%
% normalizzazioni non simmetriche
\COMMENTO{
\begin{align}
C_{(N,1)}&= 1
\nonumber\\
C_{(N,M)}&= 
\left[ 
  \frac{\prod_{j=2}^N \Gamma(\epsilon_j) }{ \Gamma(1-\epsilon_1) }
\right]^\oh
~~~~2\le M \le N-2
\nonumber\\
C_{(N,N-1)}&= 
\left[ \prod_{j=1}^N
  \frac{ \Gamma(\epsilon_j) }{ \Gamma(1-\epsilon_j) }
\right]^\oh
\end{align}
along with the OPE normalizations
\begin{align}
\cM(\alpha,\beta)
&=
1
\nonumber\\
\cN(\alpha,\beta)
&=
\sqrt{
\frac{\Gamma(\alpha)}{\Gamma(1-\alpha) } 
\frac{\Gamma(\beta)}{\Gamma(1-\beta) } 
\frac{\Gamma(2-\alpha-\beta)}{\Gamma(\alpha+\beta-1) } 
}
\nonumber\\
&=
\sqrt{
\frac{\Gamma(\alpha)}{\Gamma(1-\alpha) } 
\frac{\Gamma(\beta)}{\Gamma(1-\beta) } 
\frac{\Gamma(\gamma)}{\Gamma(1-\gamma) } 
}
~~~~
\alpha+\beta+\gamma=2
\end{align}
}
%%%%%%%%%%%%%%%%% COMMENTO %%%%%%%%%
It is at first sight surprising that there is not symmetry among the
twist operators in the $M\ne 1, N-1$ case but this is due  to two
reasons. The first is  our
choice of using a $SL(2,\R)$ invariant formalism which singles out
some points and the second is that not all twist operators are on the
same footing since some couples of twists sum to a quantity less than
one while others to one bigger than one.
These normalization are the ``square root'' of the ones found in
\cite{Erler:1992gt} for the $N=4$ closed string case and  matches
those obtained for $N=3$ in
the magnetic brane case in \cite{Duo:2007he} and for $N=4$ case in
\cite{Anastasopoulos:2011gn}.

Let us see how we can get the previous results by
exploiting the consequences of equations of the previous
subsections such as eq.s (\ref{C-C-cM}) and  (\ref{C-C-cN}).
First we notice that we can always normalize the 2-points correlator
as chosen because the generic normalization factor 
$C_{(2,1)}(\epsilon,1-\epsilon)$ is symmetric in
the exchange $\epsilon \leftrightarrow 1-\epsilon$ hence we can
redefine the twist operators as 
$\sigma_\epsilon = \tilde \sigma_\epsilon / \sqrt{ C_{(2,1)}(\epsilon,1-\epsilon)}$.

From the reduction $(N=3,M=1)$ to $(\tilde N=2,\tilde M=1)$ with the
help of eq. (\ref{C-C-M1}) we find that 
$\cM(\alpha,\beta)=C_{(3,1)}(\alpha,\beta,\gamma)$ 
with $\alpha+\beta+\gamma=1$
has the following basic symmetries
\begin{align}
\cM(\alpha, \beta )
&=
\cM(\beta, \alpha)
=
\cM(\alpha, 1-\alpha-\beta)
\end{align}
and all the others which follow from them.

In a similar way from the  $(N=3,M=2)$ to $(\tilde N=2,\tilde M=1)$
reduction  and from eq. (\ref{C-C-MN-1}) we find
\begin{align}
\cN(\alpha, \beta )
&=
\cN(\beta, \alpha)
=
\cN(\alpha, 2-\alpha-\beta)
\end{align}

Now we can consider the  $(N=4,M=2)$ to $(\tilde N=3,\tilde M=1)$
reduction in two different ways.
Either with $(\epsilon_1, \epsilon_2, \epsilon_3, \epsilon_4)
\rightarrow (\epsilon_1, \epsilon_2, \epsilon_3+\epsilon_4-1)$
which implies
\begin{align}
C_{(4,2)}(\epsilon_1, \epsilon_2, \epsilon_3, \epsilon_4)
[ B(1-\epsilon_3, 1-\epsilon_4) B(\epsilon_2, \epsilon_3+\epsilon_4-1) ]^\oh
&=
\cN(\epsilon_3, \epsilon_4) C_{(3,1)} (\epsilon_1, \epsilon_2, \epsilon_3+\epsilon_4-1)
\end{align}
or with $(\epsilon_1, \epsilon_2, \epsilon_3, \epsilon_4)
\rightarrow (\epsilon_1, \epsilon_2 +\epsilon_3, \epsilon_4-1)$
which implies
\begin{align}
C_{(4,2)}(\epsilon_1, \epsilon_2, \epsilon_3, \epsilon_4)
[ B(1-\epsilon_3, 1-\epsilon_2) B(\epsilon_4, \epsilon_3+\epsilon_2-1) ]^\oh
&=
\cN(\epsilon_3, \epsilon_2) C_{(3,1)} (\epsilon_1,
\epsilon_2+\epsilon_3-1, \epsilon_4)
\end{align}
Now taking the ratio of the two previous equations and using the
symmetries of $\cM$ and $\cN$ we are led to the minimal ansatz
\begin{align}
\cM(\alpha,\beta)
&= k
[\Gamma(\alpha) \Gamma(\beta) \Gamma(1-\alpha-\beta)]^a
[\Gamma(1-\alpha) \Gamma(1-\beta) \Gamma(\alpha+\beta)]^b
\nonumber\\
\cN(\alpha,\beta)
&= k
[\Gamma(\alpha) \Gamma(\beta) \Gamma(2-\alpha-\beta)]^c
[\Gamma(1-\alpha) \Gamma(1-\beta) \Gamma(\alpha+\beta-1)]^d
\end{align}
which gives an overconstrained system when plugged back into the ratio
constraint whose solution is $a=-b$ and $c=-d=\oh+a$.
This solution immediately yields both  $C_{(3,2)}$  and $C_{(4,2)}$.
Imposing the symmetry $\epsilon \leftrightarrow 1-\epsilon$ then
selects $a=-\frac{1}{4}$.
It is then easy to generalize to the full expressions.
These can be checked in different limits also when we consider
$\omega_j \rightarrow \infty$  using to eq. (\ref{C-C-cM-inf}).

%\subsection{The explicit $N=4$, $M=1$ case}
%\subsection{The explicit $N=4$, $M=2$ case}

%\section{Application to}
\noindent {\large {\bf Acknowledgments}}
We thank M. Bianchi for pointing out a mistake in a figure.

%%%%%%%%%%%%%%%%%%%%%%%%%%%%%%%%%%%%%%%%%%%%%%%%%%%%%%%%%%%%%%%%%%%%%%
%%%%%%%%%%%%%%%%%%%%%%%%%%%%%%%%%%%%%%%%%%%%%%%%%%%%%%%%%%%%%%%%%%%%%%
%%%%%%%%%%%%%%%%%%%%%%%%%%%%%%%%%%%%%%%%%%%%%%%%%%%%%%%%%%%%%%%%%%%%%%

\appendix
\section{Details on rewriting the classical action.}
\label{app:KLT}
\begin{figure}[hbt]
\begin{center}
\def\svgwidth{180px}
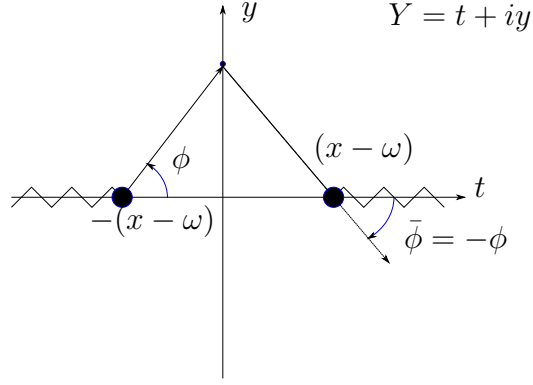
\end{center}
\vskip -0.5cm
\caption{Proper definition of angles $\phi$ and $\bar\phi$ and
  therefore of phases when $x-\omega>0$. When $x-\omega<0$ we
  substitute  $(x-\omega) \rightarrow -( x-\omega)$.}
\label{fig:KLT_phases}
\end{figure}

We want to give some details on the use of KLT technique for reducing
the integral
\begin{equation}
J^{(N)}(\alpha+n, \bar \alpha+\bar n)=
\int_{-\infty}^{+\infty} d x
\int_{-\infty}^{+\infty} d y
\prod_{j=2}^N 
(x+i y -\omega_j)^{\alpha_j+n_j}
(x-i y -\omega_j)^{\bar\alpha_j+\bar n_j}
\end{equation}
with $n_j,\bar n_j\in\Z$
to the sum of products of an holomorphic and antiholomorphic integral.
First we interpret the previous integral in $y$ as a line integral in
the complex plane $Y=t+i y$.
In the variable $Y$ the integrand has cuts in $\pm(\omega_j-x)$,
the main issue is then to properly define the phase of 
\begin{equation}
(x+ Y -\omega)^{\alpha} (x- Y -\omega)^{\bar\alpha}
=
|x+ Y -\omega|^{\alpha} |x- Y -\omega|^{\bar\alpha}
e^{i(\phi+\bar \phi)}
.
\end{equation}
The proper choice is shown in fig. (\ref{fig:KLT_phases}) and is
constrained by the request that when $Y=i y$ and $\alpha=\bar \alpha$
then $\phi+\bar \phi=0$.

\begin{figure}[hbt]
\begin{center}
\def\svgwidth{180px}
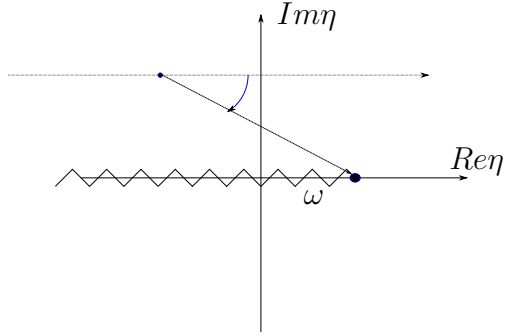
\end{center}
\vskip -0.5cm
\caption{Definition of phase in $\eta$ plane in the range $(-2\pi, 0)$.}
\label{fig:KLT_phase_eta_plane}
\end{figure}

\begin{figure}[hbt]
\begin{center}
\def\svgwidth{180px}
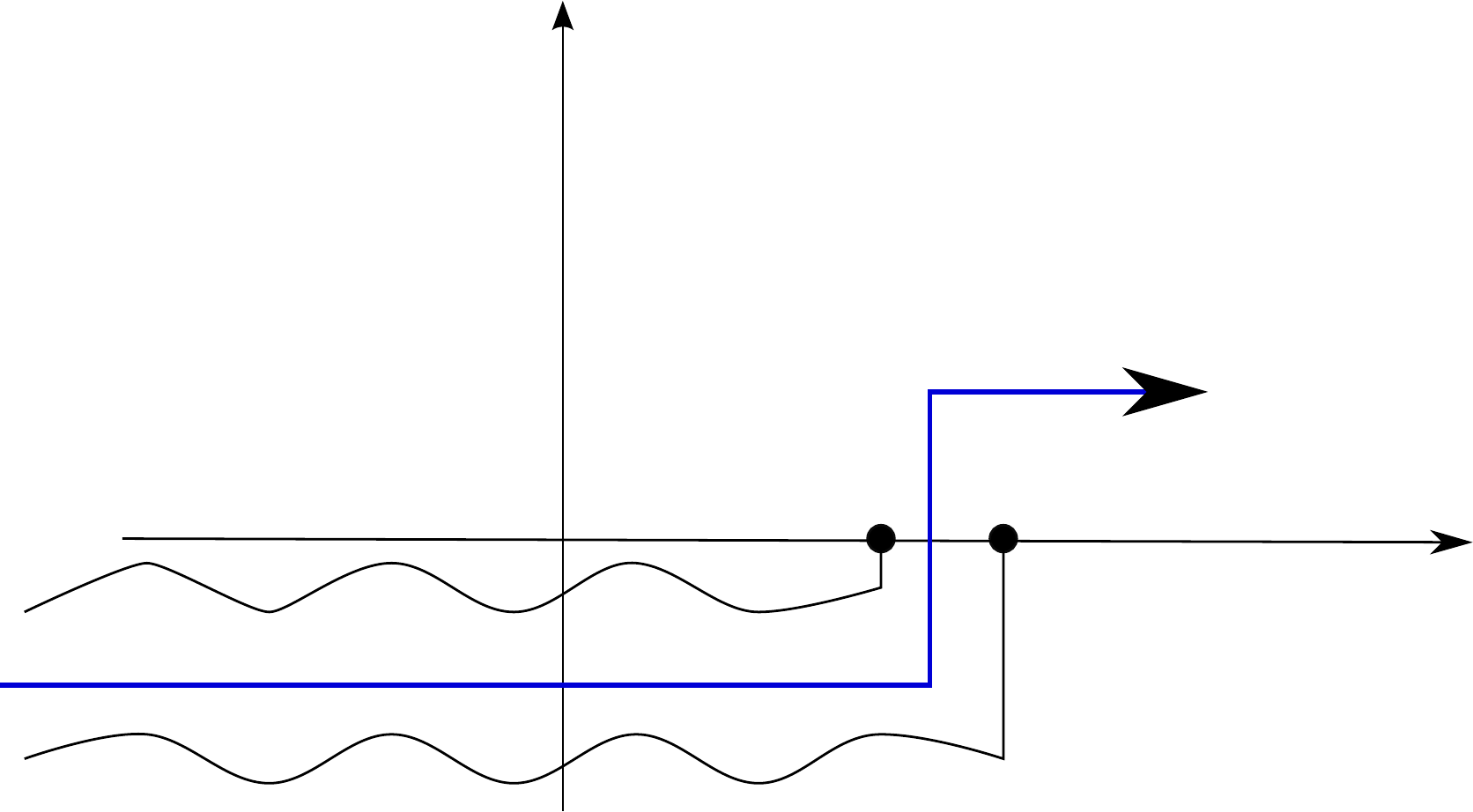
\end{center}
\vskip -0.5cm
\caption{The path $C_i$ in the complex $\eta$ plane for $\omega_{i+1}<\xi<\omega_i$.}
\label{fig:KLT_path_single_point}
\end{figure}

We can then rotate clockwise the path in $Y$ plane, change variables
as $\xi=x+t$, $\eta=x-t$
and then we can rewrite the $J^{(N)}$ integral as
\begin{align}
J^{(N)}(\alpha+n, \bar \alpha+\bar n)
=
&
-\frac{i}{2}
\int_{-\infty}^{+\infty} d \xi
\int_{-\infty}^{+\infty} d \eta
\prod_{j=2}^N 
|\xi -\omega_j|^{\alpha_j}
|\eta -\omega_j|^{\bar\alpha_j}
\nonumber\\
&~~\times
(\xi -\omega_j)^{n_j}
(\eta -\omega_j)^{\bar n_j}
\nonumber\\
&~~\times
e^{-i \pi\alpha_j ~\theta(\omega_j-\xi) ~\theta(\eta-\omega_j) }
e^{-i \pi \bar \alpha_j ~\theta(\xi-\omega_j) ~\theta(\omega_j-\eta) }
\end{align}
If $\bar \alpha=\alpha$ then we can proceed as in KLT.
We fix $\xi$ %$\omega_{i+1}< \xi< \omega_{i}$ 
and we exam the $\eta$ integral. Each factor of the integrand can then
be rewritten as
\begin{align}
|\eta -\omega_j|^{\alpha_j}
e^{-i \pi\alpha_j ~\theta( (\omega_j-\xi)(\eta-\omega_j) ) }
&=
\theta(\omega_j -\xi  )
~
[\omega_j - ( \eta+ i 0^+)]^{\alpha_j}
\nonumber\\
&+
\theta(-\omega_j +\xi  )
~
e^{+i \pi\alpha_j}
[\omega_j - ( \eta- i 0^+)]^{\alpha_j}
\end{align}
when we choose the phase in the complex $\eta$ plane as in
fig. (\ref{fig:KLT_phase_eta_plane}), obviously other choices would do
the job as well. In words this means that when
$\xi<\omega_j$ we run above the cut from $-\infty$ to $\omega_j$ in
the complex $\eta$ plane while we run below the cut when $\omega_j<\xi$.
Hence the original integral can be written as
\begin{align}
J^{(N)}(\alpha+n, \alpha+\bar n)
=
&
-\frac{i}{2}
\sum_{i=N-1}^{2}
\int_{-\omega_{i+1}}^{\omega_i} d \xi
\prod_{j=2}^N 
|\xi -\omega_j|^{\alpha_j} (\xi -\omega_j)^{n_j}
\nonumber\\
&\times
e^{i \sum_{l=i}^{2}\alpha_l}
\int_{C_i} d \eta
\prod_{j=2}^N 
(\omega_j-\eta)^{\alpha_j}
(\eta-\omega_j)^{\bar n_j}
\end{align}
where the path $C_i$ is given in fig. (\ref{fig:KLT_path_single_point}).
In particular the integrals $\int_{\omega_2}^\infty d\xi$ 
and $\int^{\omega_N}_{-\infty} d\xi$ do not contribute since the
integrals over $d \eta$ runs either above or below the cuts
and are zero because of Jordan lemma.
We can then rewrite the $C_i$ integral as an integral above (or below
depending the cases) the cuts plus a remainder. 
The final result is then
\begin{align}
J^{(N)}(\alpha+n, \alpha+\bar n)
=
&
-
%(-)^{\sum_{j=2}^N \bar n_j}
\sum_{i=2}^{N-1}
\sum_{l=i+1}^{N}
\sin\left(\pi \sum_{j=i+1}^l \alpha_j \right)
\nonumber\\
&\times
\int_{\omega_{i+1}}^{\omega_i} d \xi
\prod_{j=2}^N 
|\xi -\omega_j|^{\alpha_j} (\xi -\omega_j)^{n_j}
\nonumber\\
&\times
\int_{\omega_{l+1}}^{\omega_l} d \eta
\prod_{j=2}^N 
|\omega_j-\eta|^{\alpha_j}
(\eta-\omega_j)^{\bar n_j}
\end{align}

%%%%%%%%%%%%%%%%%%%%%%%%%%%%%%%%%%%%%%%%%%%%%%%%%%%%%%%%%%%%%%%%%%%%%%
\section{Fixing the singular part of $g(z,w)$ in a consistent way with
$N \rightarrow N-1$ reduction}
\label{app:g_sing}
Let us suppose that all coefficients $c_{n s}(\omega_j)$ depend on
$\omega_j$ ($3\le j \le N-1$) in an analytic way.
We want to show that it is then possible to fix then in a recursive
way starting from those of the $N=3, M=1$ case.
This can be done considering two limits 
$x_j \rightarrow x_N$, i.e. $\omega_j \rightarrow 0$ and
$x_j \rightarrow x_1$, i.e. $\omega_j \rightarrow \infty$.
%From the first case we get

Combining the two cases when $\epsilon_1+ \epsilon_j<1$ and
$\epsilon_j +\epsilon_N <1$ we get
\begin{align}
c^{(N,M)}_{n,s}(\omega,\epsilon)
&=
c^{(N-1,M)}_{n-1,s}(\check \omega,\check \epsilon)
-
c^{(N-1,M)}_{n,s}(\hat \omega,\hat \epsilon) \omega_j
\nonumber\\
c^{(N,M)}_{0,s}(\omega,\epsilon)
&=
-c^{(N-1,M)}_{n,s}(\hat \omega,\hat \epsilon) \omega_j
\nonumber\\
c^{(N,M)}_{N-M,s}(\omega,\epsilon)
&=
c^{(N-1,M)}_{N-M-1,s}(\check \omega, \check \epsilon)
\end{align}
when $1\le n \le N-M-1, ~~ 0\le s \le M$
and where we have defined
\begin{align}
\left\{
\begin{array}{c r}
\check \epsilon_{\check \imath}
= \epsilon_{\check \imath}
&
\check \imath =1, \dots j-1
\\
\check \epsilon_{\check \imath}
= \epsilon_{\check \imath+1}
&
\check \imath =j, \dots N-2
\\
\check \epsilon_{N-1} = \epsilon_N + \epsilon_j -\theta(\epsilon_N + \epsilon_j>1)
\end{array}
\right.
\end{align}
and
\begin{align}
\left\{
\begin{array}{c r}
\hat \epsilon_{1} = \epsilon_1 + \epsilon_j -\theta(\epsilon_1 + \epsilon_j>1)
\\
\hat \epsilon_{\hat \imath}
= \epsilon_{\hat \imath}
&
\hat \imath =2, \dots j-1
\\
\hat \epsilon_{\hat \imath}
= \epsilon_{\hat \imath+1}
&
\hat \imath =j, \dots N-1
\end{array}
\right.
\end{align}
and similar relations between $\check  \omega$ with $\omega$ and $\hat
\omega$ with $\omega$.
For example applying the previous formula to the $N=4, M=2$ case  we get
\begin{align}
g_s^{(4,2)}(z,w)
&=
\frac{1}{(z-w)^2}
\prod_{2=2}^4 \frac{(\omega_z -\omega_j)^{\epsilon_j-1}
}{
(\omega_w -\omega_j)^{\epsilon_j}
}
\Big\{
\epsilon_1 \omega_z^2 \omega_w
+(1-\epsilon_1) \omega_z \omega_w^2
\nonumber\\
&
+(1-\epsilon_1-\epsilon_2) \omega_z^2
-[\epsilon_3+\epsilon_4+(\epsilon_1+\epsilon_3)\omega_3] \omega_z \omega_w
+(1-\epsilon_2-\epsilon_4) \omega_3 \omega_w^2
\nonumber\\
&
+(1-\epsilon_4) \omega_3 \omega_z
+\epsilon_4 \omega_3 \omega_w
\Big\}
\label{g42-singular-n-into-n-1}
\end{align}

\section{Some details on the $N=4$ reduction}
\label{app:details_reductions}
As an example of the way we performed the $N\rightarrow N-1$ we give
now some details on the $N=4$ $M=2$ case, in particular 
we consider $\omega_3 \rightarrow 0$ when $\epsilon_3+\epsilon_4>1$.
Under this conditions we want to compute the behavior of
\begin{align}
\det W=
\left|
\begin{array}{c c}
W^3_1 & W^{\bar 3}_1 \\
W^3_2 & W^{\bar 3}_2 
\end{array}
\right|
\end{align}
where we have chosen $S=\bar S=\{3\}$.
The different entries of the determinant have the following limits
\begin{align}
W^3_1
&=
\int^{1}_{\omega_3} d \omega 
~(\omega-1)^{\epsilon_2-1}
~(\omega-\omega_3)^{\epsilon_3-1}
~\omega^{\epsilon_4-1}
\sim
\int^{1}_{0} d \omega 
~(\omega-1)^{\epsilon_2-1}
~\omega^{(\epsilon_3+\epsilon_4-1)-1}
\\\noindent
&= 
e^{i \pi (\epsilon_2 -1)}
B(\epsilon_2, \epsilon_3+\epsilon_4-1),
\end{align}

\begin{align}
W^3_2
&=
\int^{\omega_3}_0 d \omega 
~(\omega-1)^{\epsilon_2-1}
~(\omega-\omega_3)^{\epsilon_3-1}
~\omega^{\epsilon_4-1}
\nonumber\\
&=
\omega_3^{\epsilon_3+\epsilon_4-1}
\int^{1}_{0} d t
~(\omega_3 t -1)^{\epsilon_2-1}
~(t-1)^{\epsilon_3-1}
~t^{\epsilon_4-1}
\nonumber\\
&\sim 
\omega_3^{\epsilon_3+\epsilon_4-1}
~e^{i \pi(\epsilon_2+\epsilon_3)}
~B(\epsilon_3, \epsilon_4),
\end{align}
and
\begin{align}
W^{\bar 3}_1
&=
\int^1_{\omega_3; \omega\in H^-} d \omega 
~(\omega-1)^{-\epsilon_2}
~(\omega-\omega_3)^{-\epsilon_3}
~\omega^{-\epsilon_4}
\nonumber\\
&=
\omega_3^{1-\epsilon_3-\epsilon_4}
\int^{1/\omega_3}_{1} d t
~(\omega_3 t -1)^{-\epsilon_2}
~(t-1)^{-\epsilon_3}
~t^{-\epsilon_4}
\nonumber\\
&
\sim 
~e^{+i \pi\epsilon_2}
\omega_3^{1-\epsilon_3-\epsilon_4}
\int^{+\infty}_{1} d t
~(t-1)^{-\epsilon_3}
~t^{-\epsilon_4}
=
~e^{+i \pi\epsilon_2}
\omega_3^{1-\epsilon_3-\epsilon_4}
~B(\epsilon_3+\epsilon_4-1, 1-\epsilon_3)
\end{align}
finally the limit of the last entry can be obtained using 
again the substitution $\omega=\omega_3 t$ to be
\begin{align}
W^{\bar 3}_2
&=
\int^{\omega_3}_{0; \omega\in H^-} d \omega 
~(\omega-1)^{-\epsilon_2}
~(\omega-\omega_3)^{-\epsilon_3}
~\omega^{-\epsilon_4}
\sim
~e^{+i \pi(\epsilon_2+\epsilon_3)}
\omega_3^{1-\epsilon_3-\epsilon_4}
~B(1-\epsilon_3, 1-\epsilon_4)
\end{align}
Inserting all the previous asymptotic behaviors into the determinant
we get its $\omega_3 \rightarrow 0$, $\epsilon_3+\epsilon_4>1$ limit to be
\begin{align}
\det W
&=
\left|
\begin{array}{c c}
e^{i \pi (\epsilon_2 -1)}
B(\epsilon_2, \epsilon_3+\epsilon_4-1) 
& 
e^{i \pi \epsilon_2}
\omega_3^{1-\epsilon_3-\epsilon_4}
~B(\epsilon_3+\epsilon_4-1, 1-\epsilon_3)
\\
e^{i \pi(\epsilon_3+\epsilon_4)}
\omega_3^{\epsilon_3+\epsilon_4-1}
~B(\epsilon_3, \epsilon_4)
& 
e^{i \pi(\epsilon_2+\epsilon_3)}
\omega_3^{1-\epsilon_3-\epsilon_4}
~B(1-\epsilon_3, 1-\epsilon_4)
\end{array}
\right|
\nonumber\\
&\sim
\omega_3^{(1-\epsilon_3)+(1-\epsilon_4)-1}
~B(1-\epsilon_3, 1-\epsilon_4)
~B(\epsilon_2, 1-(1-\epsilon_3)-(1-\epsilon_4)) 
\end{align}
where it is worth noticing that we can drop the relative phases since
only one product is the leading one. This happens luckily also for all
the other computations which are needed to compute all the $N
\rightarrow N-1$ reduction.

\end{document}